\documentclass[11pt]{article}
\usepackage{verbatim,amsmath,amssymb}
\usepackage{epsfig,float,color}
\usepackage{epstopdf}
\usepackage{geometry}
\usepackage{setspace}
\usepackage{wrapfig}
\usepackage{hyperref}
\usepackage[utf8]{inputenc}
\usepackage{graphicx}
\usepackage{flushend}
\usepackage{tikz,pgfplots}
\usepackage[linesnumbered,ruled]{algorithm2e}

\usepackage[font={footnotesize}]{caption}
\usepackage[font={footnotesize}]{subcaption}
\usepackage{footnote}
\usepackage{multicol}
\usepackage{multirow}
\usepackage{enumitem}   
\usepackage{soul}
\usepackage{framed}
\usepackage[titletoc,title]{appendix}
\usepackage{bm}
\usepackage{siunitx}
\usepackage{natbib}
\usepackage[makeroom]{cancel} 

\usepackage{bigints} 

\newenvironment{rcases}
{\left.\begin{aligned}}
	{\end{aligned}\right\rbrace}

\definecolor{darkblue}{rgb}{0,0,1}
\hypersetup{pdftex=true, colorlinks=true, breaklinks=true, linkcolor=blue, menucolor=blue, citecolor=blue, urlcolor=blue}

\newcommand{\pd}[2]{\frac{\partial #1}{\partial #2}}

\usepackage{tikz}
\usetikzlibrary{positioning, arrows.meta}
\tikzset{%
	myarrow/.style = {-Stealth, shorten >=5pt}
}

\newcommand{\tr}[1]{{#1}^{\!\top}}
\newcommand{\inv}[1]{{#1}^{\text{-}1}}

\begin{document}
	
	\begin{center}
		\Large{\bf{An improved Material Mask Overlay Strategy for the desired discreteness of  pressure-loaded optimized topologies}}\\
		
	\end{center}
	
	\begin{center}

		\large{P. Kumar $^{\dagger,\star,}$\footnote{Corresponding author: pkumar@mae.iith.ac.in}, Anupam Saxena$^{\ddagger,\$}$}
		\vspace{4mm}
		
		\small{{$\dagger$}\textit{Department of Mechanical and Aerospace Engineering, Indian Institute of Technology Hyderabad, Telangana 502285, India}}\\
		\small{{$\star$}\textit{Department of Mechanical Engineering, Indian Institute of Science, Bangalore, Karnataka 560012, India}}\\
		\small{{$\ddagger$}\textit{Department of Mechanical Engineering, Indian Institute of Technology Kanpur, Uttar Pradesh 208016, India}}\\
		\small{{$\$$}\textit{Mechanical Engineering Department, RWTH Aachen University, 52062, Germany}}
			
			\vspace{4mm}
	 Published\footnote{This pdf is the personal version of an article whose final publication is available at \href{https://link.springer.com/article/10.1007/s00158-022-03401-y}{Structural and Multidisciplinary Optimization}}\,\,\,in \textit{Structural and Multidisciplinary Optimization}, 
	\href{https://link.springer.com/article/10.1007/s00158-022-03401-y}{DOI:10.1007/s00158-022-03401-y} \\
	Submitted on 01~May 2022, Revised on 02~September 2022, Accepted on 06~September 2022		
	\end{center}
	
	\vspace{1mm}
	\rule{\linewidth}{.15mm}
	{\bf Abstract:}
 This paper presents a Material Mask Overlay topology optimization approach with the improved material assignment at the element level for achieving the desired discreteness of the optimized designs for pressure-loaded problems. Hexagonal elements are employed to parametrize the design domain. Such elements provide nonsingular local connectivity; thus, checkerboard patterns and point connections inherently get subdued. Elliptical negative masks are used to find the optimized material layout. Each mask is represented via seven parameters that describe the location, shape, orientation, material dilation, and erosion variables of the mask. The latter two variables are systematically varied in conjunction with a grayscale measure constraint to achieve the solutions' sought 0-1 nature. Darcy's law with a drainage term is used to model the pressure load. The obtained pressure field is converted into the consistent nodal forces using Wachspress shape functions. Sensitivities of the objective and pressure load are evaluated using the adjoint-variable method. The efficacy and robustness of the approach are demonstrated by solving various pressure-loaded structures and pressure-driven compliant mechanisms. Compliance is minimized for loadbearing structures, whereas a multicriteria objective is minimized for mechanism designs. The boundary smoothing scheme is implemented within each optimization iteration to subdue the designs' undulated boundaries.\\
	

	\vspace{-4mm}
	\rule{\linewidth}{.15mm}
	
\section{Introduction}\label{sec:introduction}
Topology optimization (TO) is a numerical technique to find the optimized material layout within a given design domain experiencing external loads with boundary conditions by extremizing the objective subjected to a known set of constraints. Depending upon the applications, the behavior of the applied external (input) loads can be constant (design-independent) or variant (design-dependent) with the design evolution. One can find a wide range of design problems wherein design-dependent loads play crucial roles, e.g., aircraft wings and fuselage, ships, wind and snow load experiencing houses, internal and external pressure-loaded pumps and containers, pneumatically and/or hydraulically driven soft robots, etc. ~\citep{Hammer2000,kumar2020topology}. However, treatment of such loads, e.g., fluidic pressure loads in a TO setting, is challenging and involved~\citep{kumar2020topology}. This is because pressure loads' magnitude, location, and direction alter with the TO iterations. The challenges increase further as one seeks optimized,  black-and-white designs that are highly appreciated and desirable~\citep{sigmund2013topology} since the TO problems are typically relaxed to get solutions, and thus, elements with $0<\rho<1$ may exist in the optimized designs. In addition, optimized designs with gray elements cannot be realized without post-processing, which can significantly alter the performance of the fabricated designs with respect to their numerical counterparts. Further, the capability of the geometrical component-based\footnote{Typically,  geometrical component-based TO approaches require relatively lower design variables than the SIMP-based methods~\citep{kumar2015topology}.} TO approaches, for example, the Material Mask Overlay Strategy~\citep{saxena2011topology}, have not yet been explored for pressure load problems. To fill this gap, this work presents a  Material Mask Overlay topology optimization approach to solve pressure-loaded design problems wherein the target is to achieve the desired 0-1 nature of the optimized designs. The approach uses hexagonal elements to describe the design domain and masks to determine the material layout. The presented method provides an improved material assignment using the conceptualized mask dilation and erosion parameters that help reduce the number of gray elements in the optimized designs.

\begin{figure*}[h!]
	\centering
	\begin{subfigure}[t]{0.45\textwidth}
		\centering
		\includegraphics[scale = 1]{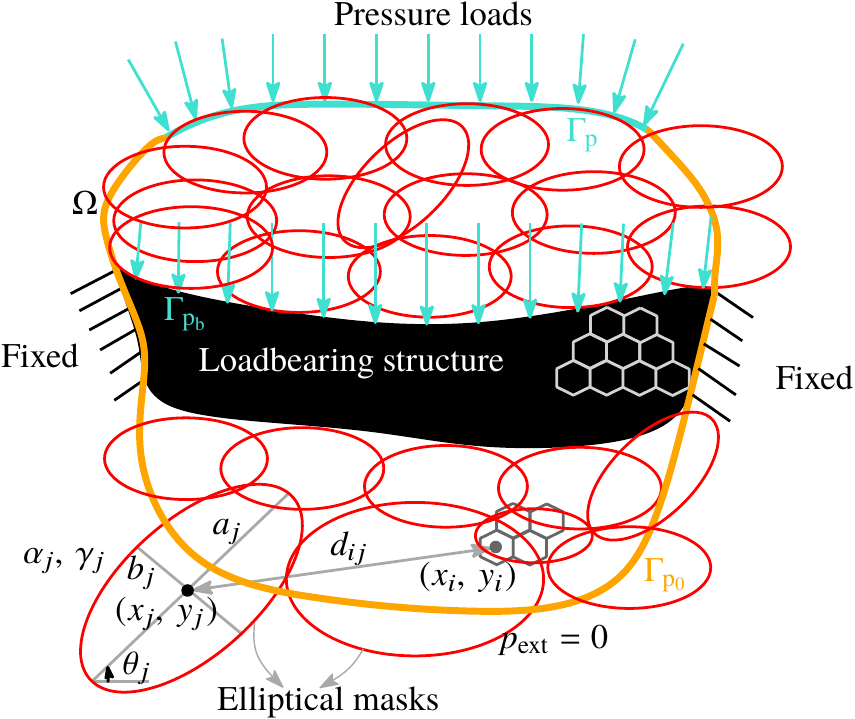}
		\caption{Loadbearing structure}
		\label{fig:schematicloadbearing}
	\end{subfigure}
	\quad
	\begin{subfigure}[t]{0.45\textwidth}
		\centering
		\includegraphics[scale = 1]{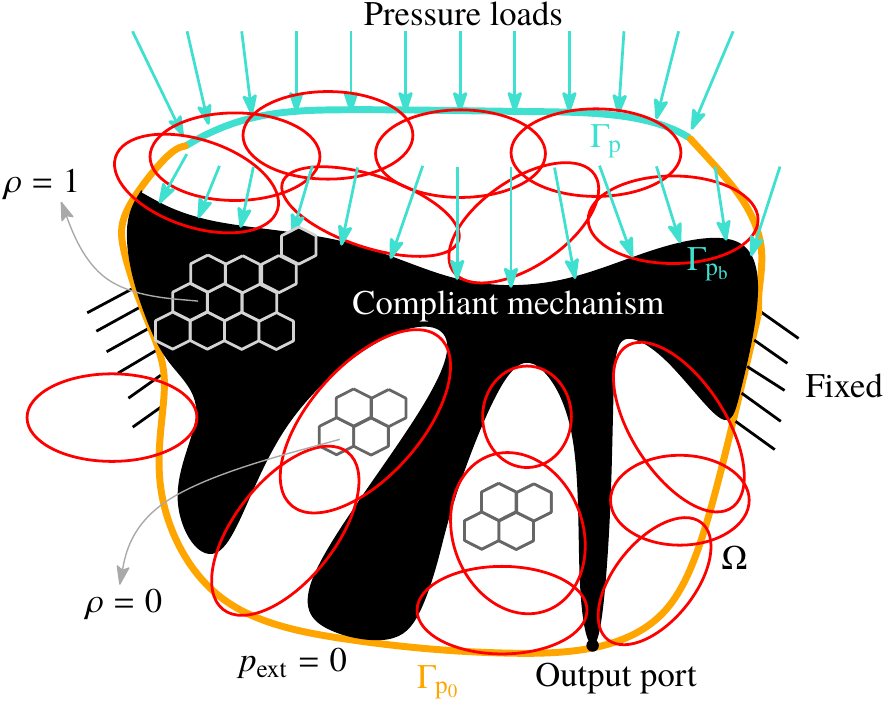}
		\caption{Compliant mechanism}
		\label{fig:schematiccompliant}
	\end{subfigure}
	\caption{Figure depicts schematic diagrams for a pressure-loaded structure and a pressure-driven CM in (\subref{fig:schematicloadbearing}) and (\subref{fig:schematiccompliant}), respectively. The design domain is denoted by $\Omega$, boundaries with finite and zero pressure loads are indicated via $\mathrm{\Gamma_\mathrm{p}}$ and $\mathrm{\Gamma_\mathrm{p_0}}$ respectively, and the final boundaries where the pressure is applied are shown by surface $\mathrm{\Gamma_\mathrm{p_b}}$. The design domain is parameterized using hexagonal elements, and negative masks are used to assign material within each finite element. Each mask is defined by $\left\{x_j,\,y_j,a_j,\,b_j,\,\theta_j,\,\alpha_j,\,\gamma_j\right\}$. The first five are \textit{geometrical variables} and latter two are termed \textit{material variables}.}	\label{fig:Schematics}
\end{figure*}

Compliant mechanisms (CMs) are monolithic designs that utilize their flexible (compliant) members to perform their tasks in response to input actuation. Such mechanisms can find various applications with/without pressure loads~\citep{kumar2019compliant,zhu2020design,kumar2021topologyDTU}. However, only a few TO approaches for pressure-driven CMs can be found~\citep{kumar2020topology,kumar2020topology3Dpressure}, and none of them present such optimized mechanisms with the desired discreteness level. With the improved material assignment for a mask,  we seek geometrical singularities free, close to black and white pressure-driven CMs and pressure-loaded structures. For the former, a multi-criteria~\citep{saxena2000optimal} objective is minimized, whereas compliance is minimized for the latter. Figure~\ref{fig:Schematics} illustrates schematic diagrams for a pressure loadbearing structure and a pressure-driven CM. One can note that the pressure loading surface moves from its initial position (surface) $\Gamma_\mathrm{p}$ to the final surface $\Gamma_\mathrm{p_b}$ (cf. Fig. \ref{fig:schematicloadbearing} and \ref{fig:schematiccompliant}) and thus, poses challenges in TO for locating and modeling. Next, we summarize existing approaches in TO for pressure-loaded designs.

The first TO approach involving pressure loads was presented by \citet{Hammer2000} for designing loadbearing structures by minimizing compliance. They used the iso-density approach to identify the pressure loading surface as TO advances. \citet{fuchs2004} employed additional variables for pressure loading boundaries. An element-based approach was presented by  \citet{zhang2008new} for locating the load surface. \citet{lee2012structural} presented an approach that does not require \textit{a prior} data of starting and ending points for pressure curves. \citet{li2018topology} proposed a regional contour tracking algorithm in conjunction with digital image processing. The approaches mentioned above either neglected the load sensitivity terms or evaluated them using the finite difference method. Load sensitivity terms are important for pressure load problems, especially while designing pressure-driven compliant mechanisms~\citep{kumar2020topology}. Level-set-based methods give implicit boundary descriptions that can be used to apply the pressure load. \citet{xia2015topology} presented a method using two zero-level functions to indicate the free and the pressure boundaries separately. Distance regularized level set evolution was used to determine structural boundary by \citet{wang2016structural}. \citet{picelli2019topology} presented the Laplace equation-based level-set TO approach to solve loadbearing structures. A bi-directional evolutionary-based TO approach for pressure load problems was presented in~\cite{picelli2015bi}. 

Instead of locating pressure loading contour explicitly, different alternate approaches were also presented. A fictitious thermal loading concept was used by \citet{Chen2001}. \citet{chen2001advances} used the method by~\cite{Chen2001} to design pressure-driven CMs.  \citet{Sigmund2007} employed the mixed finite element method with a three-phase (solid, void, and fluid) material description. A pseudo electrical potential technique was presented by \citet{Zheng2009} wherein pressure loads were directly applied upon the edges of FEs, and thus, they neglected load sensitivities. \citet{vasista2012design} employed the SIMP (Solid Isotropic Material Penalization) and MIST (Moving Isosurface Threshold) methods with the mixed displacement-pressure FE formulation. \citet{panganiban2010} used the displacement-based nonconforming FE approach that is not a trivial FE method with a three-phase material description and employed method presented by~\citet{Sigmund2007} in their approach. \citet{kumar2020topology} used  Darcy's law in association with a drainage term to design both pressure-loaded structures and pressure-driven CMs. They evaluated load sensitivity terms using the adjoint-variable method and demonstrated their effects on pressure-loaded designs. The fictitious thermal approach~\citep{Chen2001} works with a three-phase description of an element, requiring a special technique with the SIMP formulation. The Darcy law method~\citep{kumar2020topology} uses the two-state definition of an element. In addition, for the former, the pressure loads are kept constant at the boundary where they are applied for the first few iterations of optimization. However, such practices are not needed for the latter method. Further, the latter approach explicitly gives the expressions for load sensitivities, whereas the former does not. Herein, we adopt the method presented by~\cite{kumar2020topology}  for pressure-field modeling.  

To summarize, the current manuscript offers the following new aspects:
\begin{itemize}
	\item An improved  Material Masks Overlay Strategy topology optimization approach to achieve the desired close to black-and-white pressure-loaded structures and pressure-driven compliant mechanisms using honeycomb tessellation and negative circular masks
	\item Formulation of negative elliptical masks with material erosion and dilation variables to assign material density within each hexagonal element, which locally helps control the number of gray elements within the optimized design (Sec.~\ref{sec:materialmodeling})
	\item Implicitly detecting pressure loading surface using the Darcy law with hexagonal element description of the design domain in line with~\cite{kumar2020topology} (Sec.~\ref{sec:pressuremodeling})
	\item Explicitly using a grayscale measure constraint to achieve the desired discreteness level (0-1 nature) of the optimized pressure-loaded topologies while systematically varying $\{\alpha_j,\,\gamma_j\}$, i.e., material dilation and erosion variables (Sec.~\ref{sec:NumericalExamples}). 
\end{itemize}

The remainder of the paper is organized as follows. Section~\ref{sec:materialmodeling} describes density material modeling using negative elliptical masks for an FE. Section~\ref{sec:pressuremodeling} presents pressure modeling, including methodology, finite element formulation, calculation of the nodal forces, and verification problems. Topology optimization formulation, objective functions employed for the loadbearing structures and CMs under used volume and grayscale constraints, and sensitivity analysis are presented in Section~\ref{sec:optimizationPformulation}. Section~\ref{sec:NumericalExamples} reports numerical examples for structure and CM designs and pertaining discussions. Lastly, conclusions are drawn in Section~\ref{sec:closure}.
 
\section{MMOS: Material density modeling}\label{sec:materialmodeling}
In a typical TO setting with regular\footnote{In case of irregular FE discretization, nodal design variables are preferred to avoid favoring one FE over others by TO.} FE discretization descriptions, each FE is assigned a material density $\rho$. Such variables (ideally) attain either  0 or 1  values at the end of optimization and, thus, help decide the final material layout of the optimized designs. 

The Material Mask Overlay Strategy (MMOS), initially conceived in~\cite{saxena2008material} and its gradient-based version in~\cite{saxena2011topology}, is \textit{the first featured-based} TO method. The method uses masks to decide the material layout within a design domain. In a typical two-dimensional TO, a mask is constituted via a non-intersecting, analytical, or free-form closed curve~\citep{saxena2011circular,kumar2015topology,norato2018topology,singh2020topology}. A negative mask removes material from FEs over which it lays~\citep{saxena2008material,saxena2011topology}, whereas a positive mask retains material beneath it~\citep{guo2014doing,singh2020topology}. \citet{zhang2017structural} propose morphable components (MMCs) or moving morphable voids (MMVs) based method for TO. They use B-spline curves to represent the boundaries of the MMCs/MMVs. Numerous feature-mapping/geometrical components-based TO approaches exist in the current state-of-the-art of TO~\citep{wein2020review}.
\begin{figure*}[h!]
	\centering
	\begin{subfigure}{0.30\textwidth}
		\centering
		\includegraphics[scale = 0.45]{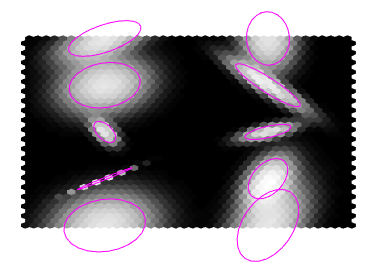}
		\caption{$\alpha_j = \gamma_j = 1$}
		\label{fig:original}
	\end{subfigure}
	\quad
	\begin{subfigure}{0.30\textwidth}
		\centering
		\includegraphics[scale = 0.45]{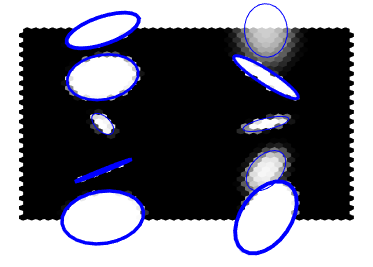} 
		\caption{Dilate: $\alpha_j\in[1,\,30]$}
		\label{fig:alphashematics}
	\end{subfigure}
	\quad
	\begin{subfigure}{0.30\textwidth}
		\centering
		\includegraphics[scale = 0.45]{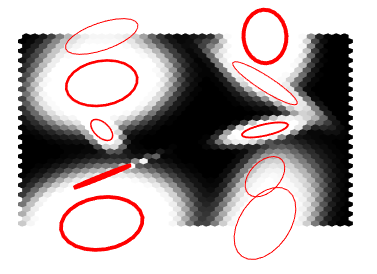}
		\caption{Erode: $\gamma_j\in[1,\,30]$}
		\label{fig:gammaschematics}
	\end{subfigure}
	\caption{Effect of $\alpha_j$ and $\gamma_j$ on the material density of FEs. (\subref{fig:original}) Design with $\alpha_j|_{j = 1,\,2,\,\cdots,\, 10} = 1$ and $\gamma_j =1$. The  design in (\subref{fig:original}) is processed by randomly varying $\alpha_j$ and $\gamma_j$ to demonstrate their effect. (\subref{fig:alphashematics}) $\alpha_j\in[1,\,30]$ are randomly varied keeping $\gamma_j = 1$. (\subref{fig:gammaschematics}). $\gamma_j\in[1,\,30]$ are randomly varied keeping $\alpha_j = 1$. The displayed line widths of masks in (\subref{fig:alphashematics}) and (\subref{fig:gammaschematics}) are as per their $\alpha_j$ and $\gamma_j$. Masks with thicker perimeter have higher material dilation and erosion variables.}	\label{fig:alphagammaschematics}
\end{figure*}

In the MMOS approach~\citep{saxena2011topology,kumar2015topology,singh2020topology}, hexagonal elements are used to parameterize the design domain, which are generated herein using $\texttt{HoneyMesher}$\footnote{Provided with \texttt{HoneyTop90} MATLAB code \citep{kumar2022honeytop90}} MATLAB code~\citep{kumar2022honeytop90}.  Edge-connectivity provided by hexagonal finite elements (FEs) subdue checkerboard patterns/point connections in optimized topologies without using additional singularity suppression schemes, e.g., filtering techniques~\citep{saxena2007honeycomb,langelaar2007use,talischi2009honeycomb,saxena2011topology,kumar2015topology,kumar2022honeytop90}. In addition, Wachspress shape functions employed to model hexagonal elements are quite rich (rational) compared to bilinear shape functions used in quadrilateral elements~\citep{talischi2009honeycomb,saxena2011topology,kumar2022honeytop90}, boundary smoothing scheme can be implemented without much difficulty as per~\cite{kumar2015topology}. Although numerical filtering can avoid checkerboards, it does not guarantee point connection-free solutions that one usually observes in compliant mechanism designs~\citep{sigmund2013topology}. Therefore, this paper employs honeycomb tessellation to parameterize the design domains, which has also not been used in the current state-of-the-art to solve pressure-loaded problems. 
 
We use negative elliptical masks wherein each mask is defined using seven variables: $x_j,\,y_j,\,a_j,\,b_j,\linebreak\theta_j,\alpha_j,\,\text{and}\,\gamma_j$. The final position, shape, size, orientation, material dilation, and erosion variables of masks determine the optimized material layout wherein the density of the $i^\text{th}$ hexagonal FE with respect to the $j^\text{th}$ elliptical mask, i.e., $\rho_{ij}$ is computed using  the logistic approximation of  Heaviside function as~\citep{singh2020topology}
\begin{equation}\label{eq:densityeachFEduetojthmasks}
	\rho_{ij}(\alpha_j) = \left[ \frac{1}{1 + \exp(-\alpha_j d_{ij})} \right],
\end{equation}
where $d_{ij}$, a Euclidean distance measure, determines  position of the centroid of the  $i^\text{th}$~FE with respect to that of the $j^\text{th}$~mask (cf. Fig.~\ref{fig:Schematics}). $\alpha_j$, material dilation variable, influences the binary nature of the solutions (Fig.~\ref{fig:alphashematics}).  Mathematically, $d_{ij}$ is evaluated as (Fig.~\ref{fig:Schematics})

\begin{equation}\label{eq:definitionofd_ij}
 d_{ij} = \left( \frac{X_{ij}}{a_j}\right)^2 + \left( \frac{Y_{ij}}{b_j}\right)^2 -1, 
\end{equation}
with,
\begin{equation}\label{eq:x_ij and y_ij}
	\begin{pmatrix}
	X_{ij}\\
	Y_{ij}
	\end{pmatrix} = \begin{bmatrix}
	\cos \theta_j & \sin \theta_j\\
	-\sin \theta_j & \cos \theta_j
	\end{bmatrix}\begin{pmatrix}
	x_i - x_j\\
	y_i - y_j
	\end{pmatrix},
\end{equation}
where $(x_i,\,y_i)$ and $(x_j,\,y_j)$ are center coordinates of the $i^\text{th}$ hexagonal FE and $j^\text{th}$ elliptical mask. $a_j\,\text{and}\,b_j$ represent the semi-major and -minor axes of the mask and $\theta_j$ is its orientation with respect to the horizontal direction. Note that the lower and upper limits for $a_j\,\text{and}\,b_j$ can be defined based on the dimension of an FE and design, and  $\theta_j\in[-\frac{\pi}{2},\,\frac{\pi}{2}]$.

In view of  $m_n$ such masks, one writes the material density of the $i^\text{th}$ FE as
\begin{equation}\label{eq:materialdensityofeachFE1}
 \rho_{i}(\alpha_j,\,\gamma_j) = \prod_{j = 1}^{m_n} \left[ \frac{1}{1 + \exp(-\alpha_j d_{ij})} \right]^{\gamma_j}, 
\end{equation} 
where $\gamma_j\in[\gamma_l,\,\gamma_u]$ and $\alpha_j\in[\alpha_l,\,\alpha_u]$.  $\alpha_j$ and $\gamma_j$ together can steer the material density of an FE towards either 0 or 1 and, thus, help ensure crisp final solutions. $\gamma_l$ and $\gamma_u$ are user-defined lower and upper bounds on $\gamma_j$. Likewise, $\alpha_l$ and $\alpha_u$ represent lower and upper limits for $\alpha_j$, which are also user-defined parameters. 

Let $\psi_j = \left\{x_j,\,y_j,a_j,\,b_j,\,\theta_j,\,\alpha_j,\,\gamma_j\right\}$. The first five are \textit{geometrical variables} and latter two are termed \textit{material variables}, positive valued, of a mask. FEs that are close to but outside mask $j$ boundary with high $\alpha_j$ tend to achieve material density $\rho\approxeq 1$ (Fig.~\ref{fig:alphashematics}). Likewise, higher $\gamma_j$ makes FEs situated just outside and/or within mask $j$ lose more material, thus making their $\rho\approxeq 0$ (Fig.~\ref{fig:gammaschematics}). Therefore, $\alpha_j$ is called \textit{ material dilation} variable, whereas $\gamma_j$ is named \textit{material erosion} variable of masks~$j$.  Fig.~\ref{fig:alphagammaschematics} demonstrates material density plots for the FEs using $2 \times 5$ masks.  The effects of $\alpha_j$ and $\gamma_j$ on the material distribution layout are indicated in Fig.~\ref{fig:alphashematics} and~\ref{fig:gammaschematics} respectively, wherein widths of the masks indicate values of respective $\alpha_j$ and $\gamma_j$.  In this work, our focus is to systematically determine $\alpha_j$ and $\gamma_j$ for each negative elliptical mask in addition to its geometrical variables, $\{x_j,\,y_j,\,a_j,\,b_j,\,\theta_j\}$, such that we achieve optimized, close to 0-1 topologies. We employ an explicit constraint on the grayscale for optimization. Note that negative masks can also be used to generate contact surfaces within them in addition to removing material if needed, for instance, while designing contact-aided designs~\citep{kumar2016synthesis,kumar2019computational,kumar2021topology}.

\section {Pressure loads modeling}\label{sec:pressuremodeling}
In conjunction with a volumetric material-dependent pressure loss, i.e., drainage term, Darcy's law is employed to relate the pressure field with material density vector $\bm{\rho}$ as per~\cite{kumar2020topology}. The associated PDE is solved using the standard finite element formulation using Wachspress shape functions \citep{wachspress1975rational,kumar2022honeytop90}. The formulation facilitates implicit detection of the pressure loading surface and conversion of the obtained pressured field into the consistent hexagonal FE nodal forces.
\subsection{Methodology}\label{subsec:methodology}
\begin{figure*}
	\centering
	\begin{subfigure}[t]{0.15\textwidth}
		\centering
		\includegraphics[scale = 0.15]{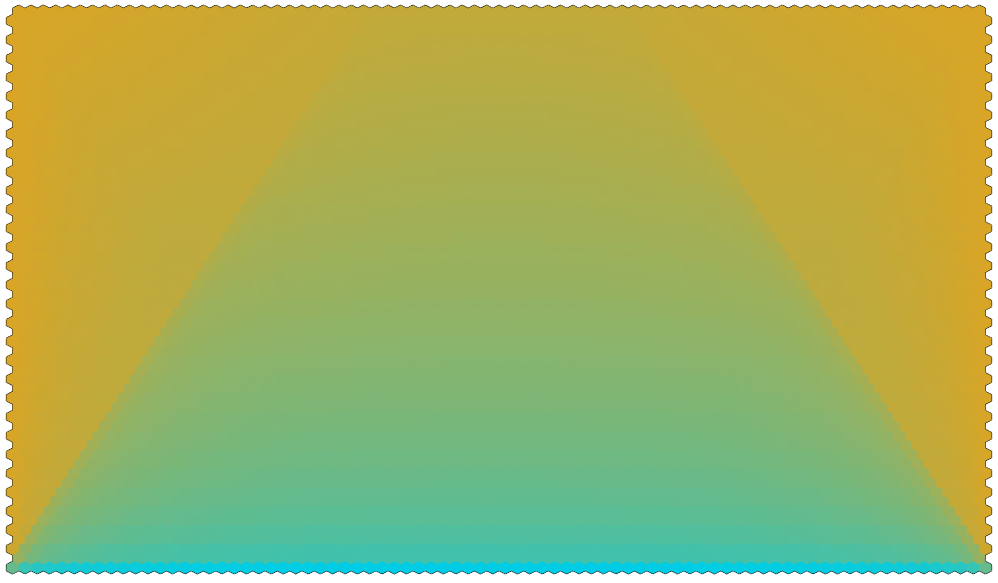}
		\caption{}
		\label{fig:PressurefieldDomainI}
	\end{subfigure}
	\quad
	\begin{subfigure}[t]{0.15\textwidth}
		\centering
		\includegraphics[scale = 0.15]{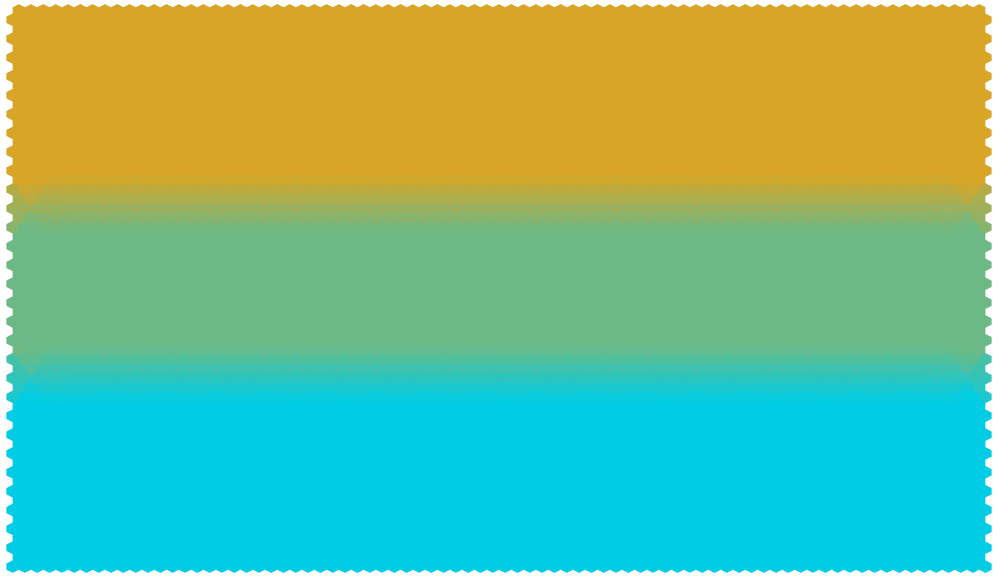}
		\caption{}
		\label{fig:PressurefieldDomainIIwithoutdrainwithoutmaterial}
	\end{subfigure}
	\quad
	\begin{subfigure}[t]{0.15\textwidth}
		\centering
		\includegraphics[scale = 0.15]{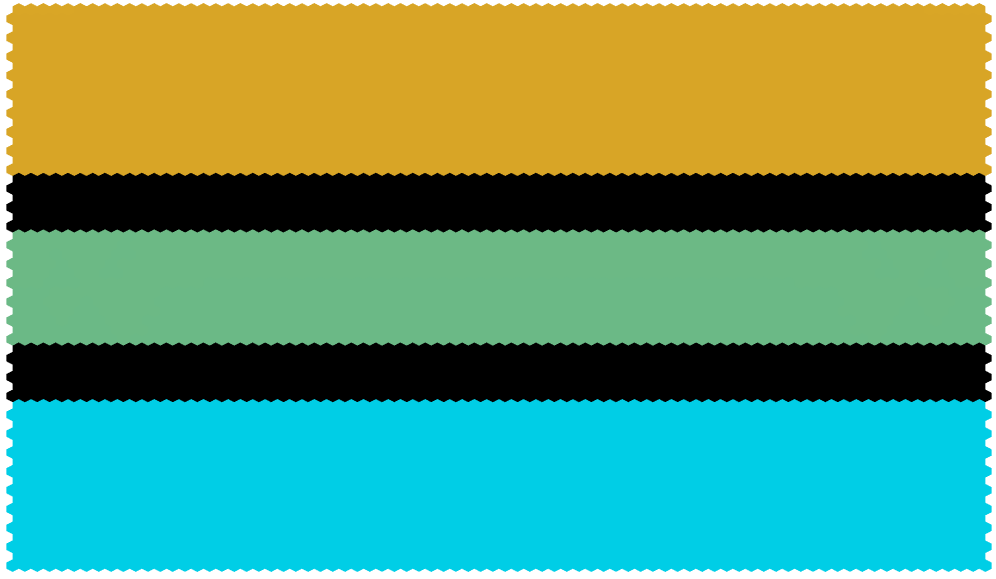}
		\caption{}
		\label{fig:PressurefieldDomainIIwithoutdrain}
	\end{subfigure}
	\quad
	\begin{subfigure}[t]{0.15\textwidth}
		\centering
		\includegraphics[scale = 0.15]{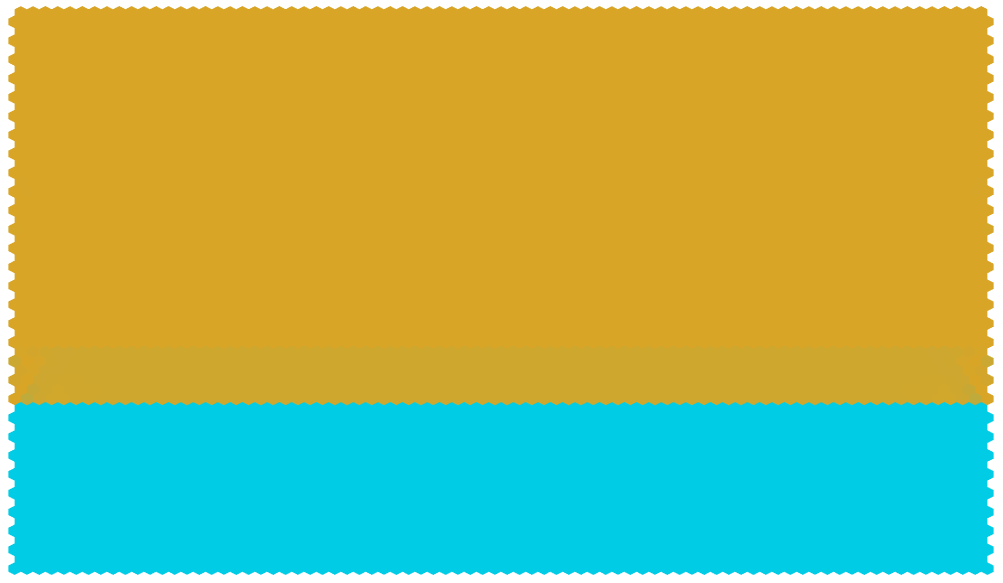}
		\caption{}
		\label{fig:PressurefieldDomainIIwithdrainwithoutmaterial}
	\end{subfigure}
	\quad
	\begin{subfigure}[t]{0.15\textwidth}
		\centering
		\includegraphics[scale = 0.15]{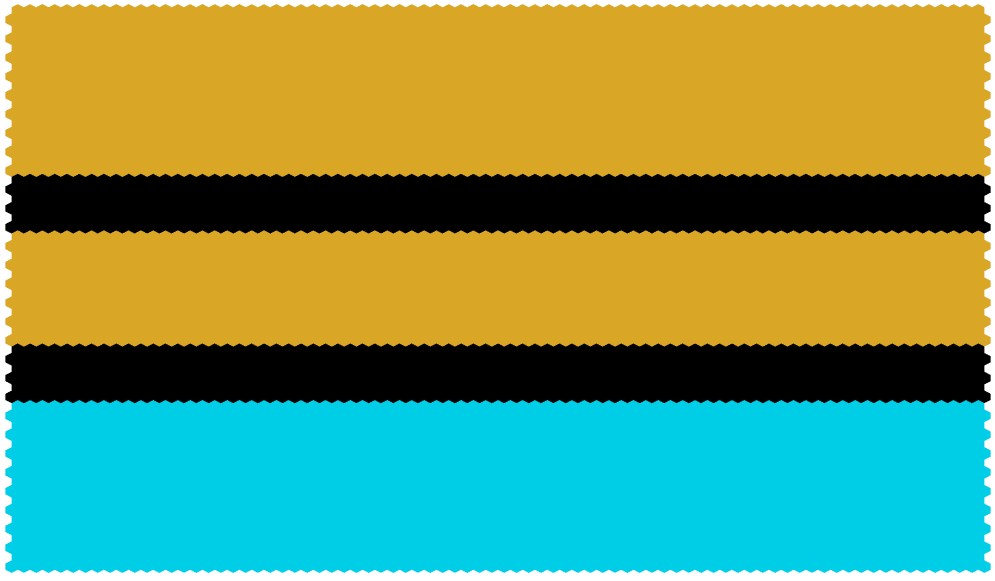}
		\caption{}
		\label{fig:PressurefieldDomainIIwithdrain}
	\end{subfigure}
	\caption{Pressure fields for DDomain~I and DDomain~II are displayed. (\subref{fig:PressurefieldDomainI})  DDomain~I pressure field (\subref{fig:PressurefieldDomainIIwithoutdrainwithoutmaterial}) DDomain~II pressure field without drainage term is plotted without solid regions (\subref{fig:PressurefieldDomainIIwithoutdrain}) DDomain~II pressure field without drainage term  with solid regions, (\subref{fig:PressurefieldDomainIIwithdrainwithoutmaterial}) DDomain~II pressure field with drainage term without solid regions and (\subref{fig:PressurefieldDomainIIwithdrain}) DDomain~II pressure field with drainage term  with solid regions. One notices that the gradient of pressure field gets confined as soon as it faces the first solid region in DDomain when using the drainage term (Fig.~\ref{fig:PressurefieldDomainIIwithdrainwithoutmaterial} and Fig.~\ref{fig:PressurefieldDomainIIwithdrain}), however the same is not noted without the drainage term (Fig.~\ref{fig:PressurefieldDomainIIwithoutdrainwithoutmaterial} and Fig.~\ref{fig:PressurefieldDomainIIwithoutdrain}).}	\label{fig:Pressurefield}
\end{figure*}

We briefly describe the Darcy law, the drainage term, and associated parameters herein. A detailed description can be found in \cite{kumar2020topology}. 
The Darcy law that helps find pressure field through a porous medium is adopted wherein the Darcy flux $\bm{q}$ depends upon the pressure gradient $\nabla p$, the fluid viscosity $\mu$ and permeability of the medium $\kappa$ as
\begin{equation}\label{eq:darcylaw}
	\bm{q} = -\frac{\kappa}{\mu}\nabla p = -K \nabla p,
\end{equation} 
where  $K$ represents the flow coefficient that refers to the ability to allow fluid to pass through a porous medium. To cater to a TO setting,  each material phase of an FE is also associated with a flow coefficient, and the actual flow coefficient of an FE is determined by performing interpolation between those associated with its solid and void material states using a smooth Heaviside projection function as
\begin{equation}\label{eq:flowcoefficient}
	K(\rho_i(\psi_j)) = K_\text{V}( 1 -(1-\epsilon)H_K(\rho_i(\psi_j),\eta_K,\beta_K)),
\end{equation}
where $\epsilon = \frac{K_\text{S}}{K_\text{V}}$ is the flow contrast~\citep{kumar2020topology3Dpressure} wherein $K_\text{S}$ and $K_\text{V}$ are the flow coefficients for solid and void  phased FEs, respectively. $H_k(\rho_i,\eta_K,\,\beta_K)$ is a smooth Heaviside projection function defined as

\begin{equation}\label{eq:Heaviside}
H_K(\rho_i(\psi_j),\eta_K,\,\beta_K) = \left(\frac{\tanh{\left(\beta_K\eta_K\right)}+\tanh{\left(\beta_K(\rho_i - \eta_K)\right)}}{\tanh{\left(\beta_K \eta_K\right)}+\tanh{\left(\beta_K(1 - \eta_K)\right)}}\right),
\end{equation}
where $\eta_K$ and $\beta_K$ help  control position of  the step and slope of $K(\rho_i(\psi_j))$ respectively. $\rho_i(\psi_j)$ is evaluated using Eq.~\eqref{eq:materialdensityofeachFE1} indicating that the defined flow coefficient $K(\rho_i(\psi_j))$ depends upon the position, shape, size, orientation, material dilation and erosion variables of the masks employed in TO. In a typical TO setting, using Darcy's law alone may fail to ensure the desire pressure field for a reasonable design as it provides pressure gradient throughout the design domain (see Fig.~\ref{fig:PressurefieldDomainI}). Therefore,  a drainage term conceptualized in \cite{kumar2020topology} and qualified in \cite{kumar2020topology3Dpressure} is employed to ensure a sharp and continuous pressure drop as soon as pressure loads encounter a solid FE while TO progresses (see Fig.~\ref{fig:PressurefieldDomainIIwithdrainwithoutmaterial} and Fig.~\ref{fig:PressurefieldDomainIIwithdrain}), i.e., drainage term becomes active when pressure loads faces solid FEs otherwise remains inactive. ${Q}_\text{drain}$ is defined as
\begin{equation}\label{Eq:Drainageterm}
{Q}_\text{drain} = -D(\rho_i(\psi_j)) (p - p_{\text{ext}}),
\end{equation}
where the pressure field and external pressure are indicated via $p$ and $p_\text{ext}$, respectively and $D(\rho_i)$ is the drainage coefficient defined using a smooth Heaviside function as
\begin{equation}\label{eq:Drainagecoefficient}
D(\rho_i(\psi_j))    = \text{D}_{\text{S}}\,H_\text{D}(\rho_i(\psi_j),\,\eta_\text{D},\,\beta_\text{D}),
\end{equation}
where $\eta_\text{D}$ and $\beta_\text{D}$ are adaptable parameters and \linebreak$H_\text{D}(\rho_i(\psi_j),\,\eta_\text{D},\,\beta_\text{D})$ is analogous to that mentioned in Eq.~\eqref{eq:Heaviside}. $\text{D}_\text{S}$ is the drainage coefficient of a solid hexagonal FE that controls the pressure-penetration depth and is determined in terms of $K_\text{S}$ as~\citep{kumar2020topology}
\begin{equation}\label{eq:hsrelation}
\text{D}_\text{s} =\left(\frac{\ln{r}}{\Delta s}\right)^2 K_\text{s},
\end{equation} 
where $r = \frac{p|_{\Delta s}}{p_\text{in}}$; $\Delta s$, a penetration parameter, is set to  width/height of a few FEs, and $p_\text{in}$ and $p|_{\Delta s}$ are input pressure and pressure at $\Delta s$, respectively. 
\subsection{Finite element formulation for pressure loading}
 The basic balance equation for Darcy's law in conjunction with  ${Q}_\text{drain}$ and incompressible fluid flow assumptions can be written as~\citep{kumar2020topology}
\begin{equation}\label{eq:balancedequation}
\nabla\cdot\bm{q} -{Q}_\mathrm{drain} = 0,
\end{equation}
In view of Eq.~\eqref{eq:darcylaw}, Eq.~\eqref{eq:balancedequation} yields 
\begin{equation}\label{eq:balancedequationpressure}
\nabla\cdot (K\nabla p) + {Q}_\mathrm{drain} = 0.
\end{equation}
The PDE in Eq.~\eqref{eq:balancedequationpressure} is solved to evaluate pressure field using the Galerkin method of finite element formulation as
\begin{equation}\label{eq:weakform1}
\sum_{i=1}^{Nel}\left(\int_{\Omega_i}\nabla\cdot (K\nabla p) G\,\text{d}V + \int_{\Omega_i}{Q}_\mathrm{drain} G \,\text{d}V \right) = 0,
\end{equation}
where $Nel$ indicates the total number of hexagonal FEs employed to describe the design domain $\Omega$, $\Omega_i|_{i = 1,\,2,\,3,\,\cdots,\,Nel}$ represent hexagonal FEs, d$V$ is the elemental volume, and $G$ is determined using the same basis functions that are employed for interpolating pressure. For a hexagonal FE
\begin{equation}\label{eq:elementalpressure}
	p = \mathbf{N}_\text{p}\mathbf{p}_l, \qquad \text{and}\qquad G = \mathbf{N}_\text{p}\mathbf{G}_l,
\end{equation}
where $\mathbf{p}_l = [p_1,\,p_2,\,p_3,\,p_4,\,p_5,\,p_6]^\text{T}$  are the hexagonal nodal pressures  and $\mathbf{N}_\text{p} = [N_1,\, N_2,\,N_3,\,N_4,\,N_5,\,N_6]$ are the Wachspress shape functions (see Appendix of \cite{kumar2022honeytop90}).  Using integration by parts, divergence theorem and  Eq.~\eqref{eq:elementalpressure}, one writes Eq.~\eqref{eq:weakform1} for element~$i$ as
\begin{figure*}
	\centering
	\begin{subfigure}{0.48\textwidth}
		\centering
		\includegraphics[scale=0.45]{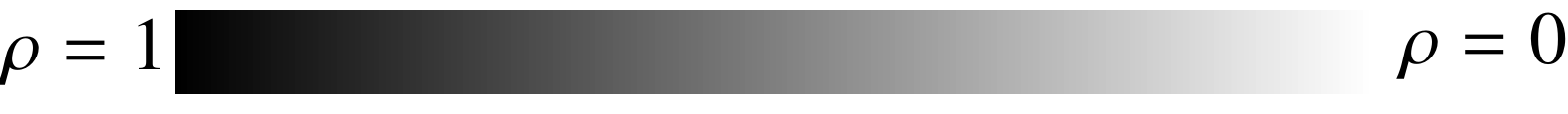}
		\caption{Material field scale}
		\label{fig:materialcolor}
	\end{subfigure}
	\begin{subfigure}{0.48\textwidth}
		\centering
		\includegraphics[scale=0.45]{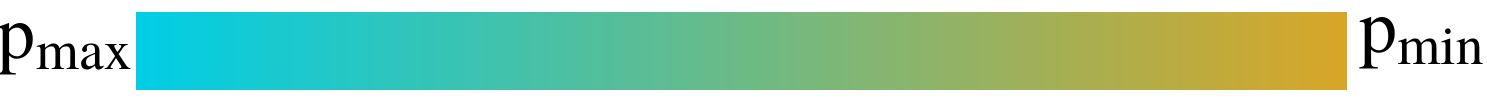}
		\caption{Pressure field scale}
		\label{fig:pressurecolor}
	\end{subfigure}
	\caption{Scales for the material density field and pressure field are displayed in (\subref{fig:materialcolor}) and  (\subref{fig:pressurecolor}), respectively, which are employed in this paper to show the optimized results and final pressure field. p$_\text{max}=\SI{1}{\bar}$ and p$_\text{min}=\SI{0}{\bar}$ are used unless otherwise stated.} \label{fig:Colorshceme}
\end{figure*}
\begin{figure*}[h!]
	\centering
	\begin{subfigure}{0.45\textwidth}
		\centering
		\includegraphics[scale = 0.5]{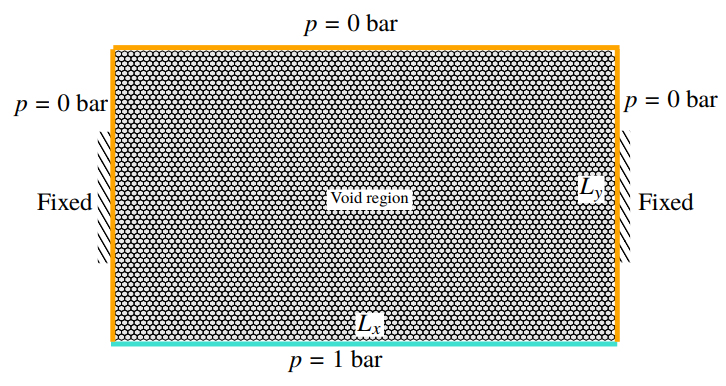}
		\caption{DDomain~I}
		\label{fig:DDomain I}
	\end{subfigure}
	\quad
	\begin{subfigure}{0.45\textwidth}
		\centering
		\includegraphics[scale = 0.5]{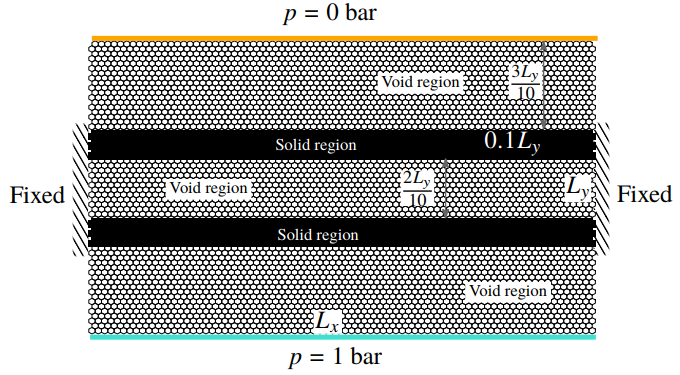}
		\caption{DDomain~II}
		\label{fig:DDomain II}
	\end{subfigure}
	\caption{DDomain~I and DDomian~II are depicted in (\subref{fig:DDomain I}) and (\subref{fig:DDomain II}), respectively. The material density of each FE in (\subref{fig:DDomain I}) is set to 0.01. $L_x= 0.2\cos(\frac{\pi}{6})\,\si{\meter},\,\text{and}\, L_y = 0.2\sin(\frac{\pi}{6})\,\si{\meter}$, designs are parameterized using $80\times 60$ FEs. DDomain~II has two solid FE (dark) layers of width $0.1L_y$ separated by $0.2L_y$. Fixed locations, pressure, and zero pressure loading edges are shown.}	\label{fig:testDomains}
\end{figure*}

\begin{equation} \label{eq:FEAformulation}
\begin{aligned}
\int_{\Omega_i}\left( K~ \tr{\mathbf{B}_\text{p}} \mathbf{B}_\text{p}   +  D ~\tr{\mathbf{N}_\text{p}} \mathbf{N}_\text{p} \right)\text{d} V~\mathbf{p}_i &=
\int_{\Omega_i}~D~\tr{\mathbf{N}_\text{p}} p_\mathrm{ext} ~~\text{d} V -
	\int_{\mathrm{\Gamma}_i}~ \tr{\mathbf{N}_\text{p}} \mathbf{q}_\mathrm{\Gamma} \cdot \mathbf{n}_i~~\text{d} A\\
	\mathbf{A}_i\mathbf{p}_i &= \mathbf{f}_i
\end{aligned}
\end{equation}
where the flux through the boundary $\mathrm{\Gamma}_i$ is represented via $\mathbf{q}_{\Gamma}$, $\mathbf{B}_\text{p} =\nabla\mathbf{N}_\text{p}$, $\mathbf{n}_i$ indicates the outward normal to the surface $\Gamma_i$, and $\text{d}A$ is the elemental area. Eq.~\eqref{eq:FEAformulation} transpires in global sense to
\begin{equation}\label{eq:fluidglobal}
\mathbf{Ap} = \mathbf{f}.
\end{equation}
The global flow matrix $\mathbf{A}$, the global pressure vector $\mathbf{p}$ and the global loading vector $\mathbf{f}$ are obtained by assembling corresponding elemental $\mathbf{A}_i$, $\mathbf{p}_i$ and $\mathbf{f}_i$, respectively. In this work, $p_\text{ext}$ and $\mathbf{q}_{\Gamma}$ are set to zero, therefore, $\mathbf{AP} = \mathbf{0}$ is solved to evaluate pressure field with given pressure loads at input locations. Each node has only one degree of freedom corresponding to pressure load; thus, it is computationally cheap to solve. The global hexagonal nodal forces recorded in $\mathbf{F}$ are determined as
\begin{equation}\label{eq:forcetransformation}
\mathbf{F} = -\mathbf{T}\mathbf{p},
\end{equation}
where $\mathbf{T}$ is a transformation matrix evaluated by assembling elemental $\mathbf{T}_i$ determined as~\citep{kumar2020topology}
\begin{equation}\label{eq:nodalforce}
\mathbf{T}_i =  -\int_{\Omega_i} \tr{\mathbf{N}_\mathbf{u}} \mathbf{B}_\text{p} \text{ d} V,
\end{equation}
where $\mathbf{N}_\mathbf{u} = \left[N_1\mathbf{I},\,N_2\mathbf{I},\,N_3\mathbf{I},\,N_4\mathbf{I},\,N_5\mathbf{I},\,N_6\mathbf{I}\right]$, $N_l|_{l= 1,\,2,\,\cdots,\,6}$ are Wachspress shape functions and $\mathbf{I}$ is the identity matrix in $\mathcal{R}^2$. Integrations in Eqs.~\eqref{eq:FEAformulation} and \eqref{eq:nodalforce} are evaluated using the quadrature rule  mentioned in \cite{kumar2022honeytop90}.  To summarize, Eq.~\ref{eq:fluidglobal} and Eq.~\ref{eq:forcetransformation} are solved to determine respectively the pressure field and the consistent nodal forces, which are further used to evaluate the state variable vector $\mathbf{u}$ (Sec.~4).

\subsection{Pressure modeling verification} 
To demonstrate the employed pressure modeling scheme (Sec.~\ref{sec:pressuremodeling}) with hexagonal FEs simulated using Wachspress shape functions, we consider two design domains: DDomain~I (Fig.~\ref{fig:DDomain I}) and DDomain~II (Fig.~\ref{fig:DDomain II}) with respective pressure and structural boundary conditions (Fig.~\ref{fig:testDomains}). Each hexagonal FE of DDomain~I is assigned low material density $\rho = 0.01$. DDomain~II is with two solid material regions which are introduced to illustrate the behavior of the drainage term (Eq.~\ref{eq:Drainagecoefficient}). $\rho=0.01$ is assigned to each FE associated with the remaining domain of DDomain~II. The bottom edge of DDomain~I experiences pressure load, whereas its remaining edges are kept at zero pressure load. In DDomain~II, the top and bottom edges experience zero and full pressure loads, respectively. Other specifications are indicated in Table~\ref{Table:T1}. The employed scales for material density field and pressure field in this paper are plotted in Fig.~\ref{fig:materialcolor} and Fig.~\ref{fig:pressurecolor}, respectively. Plane-stress conditions are considered for all other design problems solved in this paper.

The pressure field obtained by solving  Eq.~\eqref{eq:FEAformulation}  is depicted in Fig.~\ref{fig:PressurefieldDomainI} for  DDomain~I, and that for DDomain~II without and with drainage terms are plotted in Fig.~\ref{fig:PressurefieldDomainIIwithoutdrainwithoutmaterial} and Fig.~\ref{fig:PressurefieldDomainIIwithdrainwithoutmaterial}, respectively. Fig.~\ref{fig:PressurefieldDomainIIwithoutdrain} and Fig.~\ref{fig:PressurefieldDomainIIwithdrain} indicate the pressure field with solid material layers. One notices that a pressure gradient exists within DDomain~I, which is expected per Darcy's law. The material density for each FE in DDomain~I is kept low ($\rho=0.01$); consequently, the drainage term (Eq.~\ref{eq:Drainagecoefficient}) remains always inactive. One also notices without drainage term, the obtained pressure field is not realistic for DDomain~II (Fig.~\ref{fig:PressurefieldDomainIIwithoutdrainwithoutmaterial} and Fig.~\ref{fig:PressurefieldDomainIIwithoutdrain}), whereas Fig.~\ref{fig:PressurefieldDomainIIwithdrainwithoutmaterial} and Fig.~\ref{fig:PressurefieldDomainIIwithdrain} indicate the desired pressure field. Thus, the conceptualized drainage term is indeed essential. The obtained pressure fields of DDomain~II with and without drainage term are converted into the nodal force using Eq.~\ref{eq:forcetransformation}.

\section{Optimization problem formulation}\label{sec:optimizationPformulation}
This section presents the optimization problem formulation and  sensitivity analysis of the objectives employed for designing pressure loadbearing structures and pressure-actuated CMs. 

Let $\bm{\psi}$ be the design vector that stacks seven variables ($x_j,\,y_j,\,a_j,\,b_j,\,\theta_j,\,\alpha_j,\,\gamma_j|_{j=1,\,\cdots,\,m_n}$) which define each mask. The material densities of all FEs stacked in a vector, as per Eq.~\ref{eq:materialdensityofeachFE1} a function of $\bm{\psi}$, is denoted via $\bm{\rho}$. The following optimization problems are solved:

\begin{equation}\label{eq:Optimizationequation}
\begin{rcases}
\begin{split}
&\quad \, \overbrace{\underset{\bm{\rho(\bm{\psi})}}{\text{min}}\,\,\left(f_\text{s}^0=2SE\right)}^{\text{Structures}}\,\,\,\, \,\, \overbrace{\underset{\bm{\rho(\bm{\psi})}}{\text{min}}\,\,\left(f_\text{CM}^0=-\chi\frac{MSE}{SE}\right)}^{\text{Compliant Mechanisms}}\\
&\text{s.t.} \quad\qquad\,\qquad \qquad\qquad\, \qquad\mathbf{K(\bm{\rho(\bm{\psi})})v = F_\mathrm{d}}\\
& \, \qquad\,\qquad\, \mathbf{A(\bm{\rho(\bm{\psi})})p} = \mathbf{0 }\\
& \,\qquad\, \qquad\mathbf{K(\bm{\rho(\bm{\psi})})u = F} = -\mathbf{T(\bm{\rho(\bm{\psi})}) p}\\
&  \,\qquad\, \qquad\text{g}_1=\frac{ V(\bm{\rho(\bm{\psi})})}{V^*}-1\le 0\\
&  \,\qquad\, \qquad\text{g}_2=GS_\text{I} =  \frac{\sum_{i=1}^{Nel}4\rho_i(1-\rho_i)}{Nel}\le \delta\\
&  \,\qquad\, \qquad\bm{\psi} = \left[x_j,\,y_j,\,a_j,\,b_j,\,\theta_j,\,\alpha_j,\,\gamma_j\right]_{j= 1,\,\cdots,\,m_n}\\
& \, \qquad\, \qquad\bm{\psi}_\text{min}\le\bm{\psi}\le \bm{\psi}_\text{max}\\ 
\end{split}
\end{rcases},
\end{equation}
where $SE=\tr{\mathbf{u}}\mathbf{K}\mathbf{u}$ and $MSE = \tr{\mathbf{v}}\mathbf{K}\mathbf{u}$ represent strain energy and mutual-strain energy, respectively. $\mathbf{u}$ determined using $\mathbf{K(\bm{\rho(\bm{\psi})})u = F}$, and $\mathbf{v}$ evaluated employing\footnote{$\mathbf{K(\bm{\rho(\bm{\psi})})v = F_\text{d}}$ is solved only while designing CMs.} $\mathbf{K(\bm{\rho(\bm{\psi})})v = F_\text{d}}$, are the global displacement vectors corresponding to the forces $\mathbf{F}$ and $\mathbf{F}_\text{d}$. $\mathbf{F}_\text{d}$ is a dummy unit force applied in the direction of the desired output deformation of the CMs, whereas $\mathbf{F}$ is evaluated using Eq.~\eqref{eq:forcetransformation}. $\mathbf{K}$ is the global stiffness matrix of the design domain evaluated by assembling  elemental stiffness $\mathbf{k}_i= \left[E_\text{min} + \rho_i (\alpha_j,\,\gamma_j)(E_1-E_\text{min})\right]\mathbf{k}_0$,\, $E_1$ and $E_\text{min}$ are the Young's moduli of a solid and void FE, respectively, $\rho_i (\alpha_j,\,\gamma_j)$ is the material density of the $i^\text{th}$ FE (Eq.~\ref{eq:materialdensityofeachFE1}) and $\mathbf{k}_0$ is the elemental stiffness matrix for a solid FE at unit elastic modulus. Further, $\chi$, a consistent scaling factor, is primarily employed to adjust sensitivities of the objective pertaining to CMs~\citep{saxena2000optimal}. $g_1$, an inequality constraint, guides to achieve the optimized design with the permitted resource volume. $V(\bm{\rho(\bm{\psi})})$ and $V^*$ indicate the current and permitted volumes of the design domain, respectively.  $g_2$ is the grayscale indicator \citep{sigmund2007morphology} constraint, and $\delta$ is a user-defined (very) small positive number. This constraint is applied to motivate the optimization process towards  0-1 solutions. $\alpha_j$ (material dilation) and $\gamma_j$ (material erosion) are the additional design variables as mentioned in Sec~\ref{sec:materialmodeling}. $\mathbf{A}$, $\mathbf{T}$, and $\mathbf{p}$ represent the global flow matrix, global transformation matrix, and global pressure loads vector, respectively.  $\bm{\psi}_\text{min}$ and $\bm{\psi}_\text{max}$ are the lower and upper limits on the design vector $\bm{\psi}$ respectively.

\subsection{Sensitivity analysis}
The Method of Moving Asymptotes (MMA, cf.~\citet{svanberg1987method}), a gradient-based optimizer, is used herein to solve the optimization problems. Therefore one requires to have sensitivities of the objective(s) and constraint(s) with respect to design vector $\bm{\psi}$ for the optimization. One notes (Eq.~\ref{eq:Optimizationequation}), objectives ($SE$ and $-\frac{MSE}{SE}$) and constraints are function of the material density vector~$\bm{\rho}$ and that depends upon~$\bm{\psi}$; therefore, a chain rule is employed for determining the sensitivities, which is described below.

 Say, $\psi_j^1$ represents any one of the $\left\{x_j,\,y_j,\,a_j,\,b_j,\,\theta_j\right\}$ and $\psi_j$ = $\left\{\psi_j^1,\,\alpha_j,\,\gamma_j\right\}$. Using Eq.~\eqref{eq:materialdensityofeachFE1}, derivative of $\rho_i$ with respect to $\psi_j^1$ can be evaluated as
\begin{equation}\label{eq:rho over psi}
 \frac{\partial \rho_{i} (\alpha_j, \gamma_j) }{\partial \psi_{j}^1} = \gamma_j \alpha_j \rho_{i} (\alpha_j, \gamma_j)  \left[  1 -  \frac{1}{1 + \exp(-\alpha_j d_{ij})}   \right]  \left[\frac{\partial d_{ij}}{\partial \psi_{j}^1}  \right], 
\end{equation}
where $\left[\frac{\partial d_{ij}}{\partial \psi_{j}^1}  \right]$ are evaluated using Eq.~\ref{eq:definitionofd_ij} and Eq.~\ref{eq:x_ij and y_ij} as
 \begin{figure*}
	\centering
	\begin{subfigure}[t]{0.45\textwidth}
		\centering
		\includegraphics[scale=1]{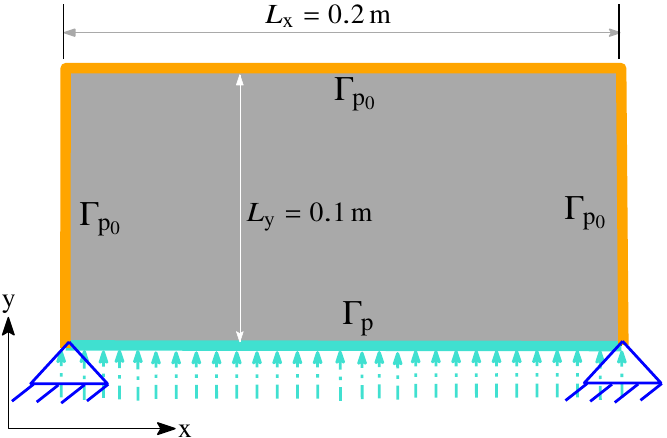}
		\caption{}
		\label{fig:lid}
	\end{subfigure}
	\quad
	\begin{subfigure}[t]{0.45\textwidth}
		\centering
		\includegraphics[scale=1]{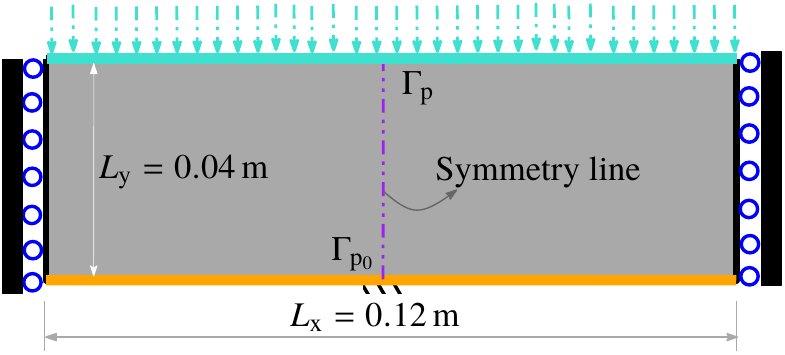}
		\caption{}
		\label{fig:piston}
	\end{subfigure}
	\caption{Design domains for Internally pressurized  and piston loadbearing structures in (\subref{fig:lid}) and (\subref{fig:piston}) respectively. Displacement and pressure boundary conditions are also depicted. $\Gamma_\text{p}$ and $\Gamma_\mathrm{p_0}$ indicate edges with $\SI{1}{\bar}$ and $\SI{0}{\bar}$ pressure loads respectively.} \label{fig:DesigndomainsStrucutres}
\end{figure*}
	\begin{figure*}
	\centering
	\begin{subfigure}[t]{0.45\textwidth}
		\centering
		\includegraphics[scale=1]{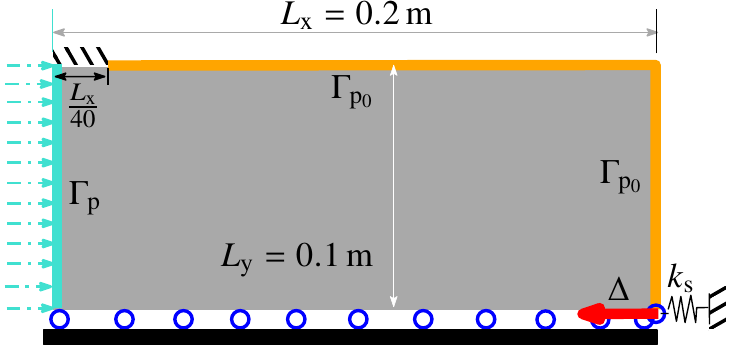}
		\caption{}
		\label{fig:inverter}
	\end{subfigure}
	\qquad
	\begin{subfigure}[t]{0.45\textwidth}
		\centering
		\includegraphics[scale=1]{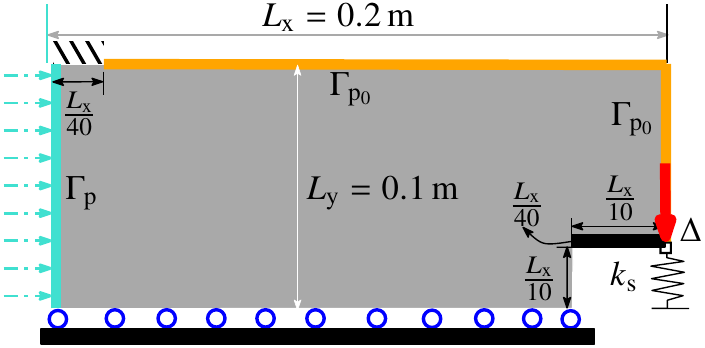}
		\caption{}
		\label{fig:gripper}
	\end{subfigure}
	\caption{(\subref{fig:inverter}) Inverter mechanism design domain and (\subref{fig:gripper}) Gripper mechanism design domain. Displacement and pressure boundary conditions are also depicted. $\Gamma_\text{p}$ and $\Gamma_\mathrm{p_0}$ indicate edges with $\SI{1}{\bar}$ and $\SI{0}{\bar}$ pressure loads respectively.} \label{fig:DesigndomainsCMs}
\end{figure*}
\begin{equation}
\begin{rcases}
\begin{split}
&&\frac{\partial d_{ij}}{\partial x_{j}} =  2\left[-\left( \frac{X_{ij}}{a_j}\right)\left( \frac{\cos \theta_j}{a_j}\right) + \left( \frac{Y_{ij}}{b_j}\right)\left( \frac{\sin \theta_j}{b_j}\right)\right]\\
&&\frac{\partial d_{ij}}{\partial y_{j}} = -2 \left[\left( \frac{X_{ij}}{a_j}\right)\left( \frac{\sin \theta_j}{a_j}\right) +  \left( \frac{Y_{ij}}{b_j}\right)\left( \frac{\cos \theta_j}{b_j}\right)\right] \\
&&\frac{\partial d_{ij}}{\partial a_{j}} = - 2 \frac{X_{ij}^2}{a_j^3},\, \frac{\partial d_{ij}}{\partial b_{j}} = - 2 \frac{Y_{ij}^2}{b_j^3}\\
&&\frac{\partial d_{ij}}{\partial \theta_{j}} = 2 \left(\frac{X_{ij}Y_{ij}}{a_j^2} - \frac{X_{ij}Y_{ij}}{b_j^2}\right)\\
\end{split}
\end{rcases}.
\end{equation}
Derivatives of  $\rho_{i} (\alpha_j, \gamma_j)$ with variables $\alpha_j$ and $\gamma_j$ can be found as
\begin{equation}\label{eq:rho over alpha}
\frac{\partial \rho_{i} (\alpha_j, \gamma_j) }{\partial \alpha_{j}} = \gamma_j d_{ij} \rho_{i} (\alpha_j, \gamma_j)  \left[  1 -  \frac{1}{1 + \exp(-\alpha_j d_{ij})}   \right], 
\end{equation}

\begin{equation}\label{eq:rho over eta}
\frac{\partial \rho_{i} (\alpha_j, \gamma_j) }{\partial \gamma_{j}} =  \rho_{i} (\alpha_j, \gamma_j)  \log\left(\frac{1}{1 + \exp(-\alpha_j d_{ij})}   \right).
\end{equation}
Therefore, $\frac{\partial \rho_{i}}{\partial \psi_{j}} = \left[\left[\frac{\partial \rho_{i}}{\partial \psi_{j}^1}\right]^\top  \left[\frac{\partial \rho_{i} (\alpha_j, \gamma_j) }{\partial \alpha_{j}}\right]^\top \left[\frac{\partial \rho_{i} (\alpha_j, \gamma_j) }{\partial \eta_{j}}\right]^\top\right]$.
The adjoint-variable method is used to evaluate sensitivities of the objectives with respect to the material density vector~$\bm{\rho}$. One writes the following overall performance functions $\mathcal{L}_\text{s}$ for loadbearing structures as
\begin{equation}\label{eq:performance_structure}
\mathcal{L}_\text{s} = f^0_\text{s} + \tr{\bm{\lambda}}_{\text{s}_1} \left(\mathbf{Ku +{T p}}\right) + \tr{\bm{\lambda}}_{\text{s}_2} ( \mathbf{Ap}),
\end{equation}
where $\bm{\lambda}_{\text{s}_1}$ and $\bm{\lambda}_{\text{s}_2}$ are the Lagrange multiplier vectors. Likewise, the performance function $\mathcal{L}_\text{CM}$ for CMs is
\begin{equation}\label{eq:performance_CM}
\mathcal{L}_\text{CM} = f^0_\text{CM} + \tr{\bm{\lambda}}_{\text{CM}_1} \left(\mathbf{Ku +{T p}}\right) + \tr{\bm{\lambda}}_{\text{CM}_2} (\mathbf{Ap}) + \tr{\bm{\lambda}}_{\text{CM}_3} (\mathbf{Kv-F_\mathrm{d}}),
\end{equation}
where $\bm{\lambda}_{\text{CM}_1}$,\,$\bm{\lambda}_{\text{CM}_2}$ and $\bm{\lambda}_{\text{CM}_3}$ are the Lagrange multiplier vectors. These multipliers can be determined as~\citep{kumar2020topology}

\begin{table*}
	\caption{Details of parameters used}\label{Table:T1}
	\centering
	\begin{tabular}{c c c c c}
		\hline \hline
		\textbf{Nomenclature}  & \textbf{Symbol} & \textbf{Value} &  &  \rule{0pt}{3ex}\\ \hline 
		\multicolumn{3}{c}{\textit{Masks parameters}}      &  & \rule{0pt}{3ex} \\ \hline
		No. of Masks in $x-$direction & $N_\text{mx}$                     & $20$                                 &  & \rule{0pt}{3ex} \\ 
		No. of masks in $y-$direction         & $N_\text{my}$             & $10$                                 &  &  \rule{0pt}{3ex}\\ 
		Mask radius parameter               & $mR$                  & $30\times$edge-length of an FE                                &  &  \rule{0pt}{3ex}\\ 
		Lower bound factor for the axes of a mask  & $f_l$               & $0.001 \times mR$                    &  &  \rule{0pt}{3ex} \\ 
		Upper bound factor for the axes of a mask & $f_u$               & $1 \times mR$                       &  &  \rule{0pt}{3ex} \\ 
		lower bounds for $\alpha_j$ and $\gamma_j$ & --               & $1,\,1$                       &  &  \rule{0pt}{3ex} \\ 
		Upper bounds  for $\alpha_j$ and $\gamma_j$ & --              & $30$,\,$30$                       &  &  \rule{0pt}{3ex} \\ \hline
		\multicolumn{3}{c}{\textit{Material Parameters}}           &  &  \rule{0pt}{3ex}\\\hline
		Young's modulus of a solid FE       & $E_1$               &$\SI{1e7}{\newton\per\square\meter}$ &  &  \rule{0pt}{3ex}\\ 
		Young's modulus of a void FE        & $E_\text{min}$               & $E_1\times 10^{-6} $                 &  &  \rule{0pt}{3ex}\\ 
		SIMP penalty parameter              & $\zeta$        & $1$                                        &  &  \rule{0pt}{3ex}\\  \hline
		\multicolumn{3}{c}{\textit{Pressure load parameters}}             &  &  \rule{0pt}{3ex}\\ \hline
		Input pressure load &$p_\mathrm{in}$ & $\SI{1e5}{\newton\per\square\meter}$ \rule{0pt}{3ex}& & \rule{0pt}{3ex}\\
		$K(\rho)$ step location             &               $\eta_k$ 	& $0.3$	                       &  &  \rule{0pt}{3ex}\\ 
		$K(\rho)$ slope at step             &            	$\beta_k$	& $10$                       &  &  \rule{0pt}{3ex}\\ 
		$D(\rho)$ step location             &                	$\eta_h$ 	& $0.3$                         &  &  \rule{0pt}{3ex}\\ 
		$D(\rho)$ slope at step             &                    $\beta_h$	& $10$                           &  &  \rule{0pt}{3ex}\\ 
		Flow coefficient of a void FE	&   $K_\mathrm{v}$  & $\SI{1}{\meter\tothe{4}\per\newton\per\second}$& & \rule{0pt}{3ex}\\
		Flow coefficient of a solid FE &   $K_\mathrm{s} $  & $K_\mathrm{v}\times\SI{e-7}{\meter\tothe{4}\per\newton\per\second}$ & &\rule{0pt}{3ex}\\
		Drainage from solid	&   $D_\mathrm{s} $  & $\left(\frac{\ln{r}}{\Delta s}\right)^2 K_\mathrm{s}$ & &\rule{0pt}{3ex} \\
		Remainder of input pressure at $\Delta s$ &r& 0.1& & \rule{0pt}{3ex}\\\hline\hline
	\end{tabular}
\end{table*}

\begin{equation}\label{eq:lagrangemultipliers_strctures}
\begin{rcases}
\tr{\bm{\lambda}}_{\text{s}_1}  = -\pd{f_s^0}{\mathbf{u}} \inv{\mathbf{K}} = -2\tr{\mathbf{u}}\\
\tr{\bm{\lambda}}_{\text{s}_2}   = -\tr{\bm{\lambda}}_{\text{s}_1} \mathbf{T}\inv{\mathbf{A}} = 2\tr{\mathbf{u}} \mathbf{T}\inv{\mathbf{A}} \\
\end{rcases},
\end{equation}
and
\begin{equation}\label{eq:lagrangemultiplierCM}
\begin{rcases}
\tr{\bm{\lambda}}_{\text{CM}_1}  =  -\pd{f_\text{CM}^0}{\mathbf{u}} \inv{\mathbf{K}} = \chi\left(\frac{\tr{\mathbf{v}}}{SE} - \tr{\mathbf{u}}\frac{MSE}{(SE)^2}\right)\\
\tr{\bm{\lambda}}_{\text{CM}_2} =-\tr{\bm{\lambda}}_{\text{CM}_1} \mathbf{T}\inv{\mathbf{A}}  = -\chi\left(\frac{\tr{\mathbf{v}}}{SE} - \tr{\mathbf{u}}\frac{MSE}{(SE)^2}\right)\mathbf{T}\inv{\mathbf{A}}\\ 
\tr{\bm{\lambda}}_{\text{CM}_3} =  -\pd{f_\text{CM}^0}{\mathbf{v}} \inv{\mathbf{K}}=\chi\frac{\tr{\mathbf{u}}}{SE}\\
\end{rcases}.
\end{equation}
Using Eqs.~\eqref{eq:performance_structure}, \eqref{eq:performance_CM}, \eqref{eq:lagrangemultipliers_strctures} and \eqref{eq:lagrangemultiplierCM}, sensitivities of the objective functions with respect to $\bm{\rho}$ can be written as

\begin{equation}\label{eq:objectivesensitivityrho}
\begin{split}
\frac{\text{ d} f_\text{s}^0}{\text{ d}\bm{\rho}} &= \pd{f_\text{s}^0}{\bm{\rho}} + \tr{\bm{\lambda}}_{\text{s}_1}\pd{\mathbf{K}}{\bm{\rho}}\mathbf{u} + \tr{\bm{\lambda}}_{\text{s}_2}\pd{\mathbf{A}}{\bm{\rho}}\mathbf{p}\\
&=-\tr{\mathbf{u}}\pd{\mathbf{K}}{\bm{\rho}}\mathbf{u} + \underbrace{2\tr{\mathbf{u}} \mathbf{T}\inv{\mathbf{A}}\pd{\mathbf{A}}{\bm{\rho}}\mathbf{p}}_{\text{Load sensitivities}}\\
\text{and}\\
\frac{\text{ d} f_\text{CM}^0}{\text{ d}\bm{\rho}} &= \pd{f_\text{CM}^0}{\bm{\rho}} + \tr{\bm{\lambda}}_{\text{CM}_1}\pd{\mathbf{K}}{\bm{\rho}}\mathbf{u} + \tr{\bm{\lambda}}_{\text{CM}_2}\pd{\mathbf{A}}{\bm{\rho}}\mathbf{p} +  \tr{\bm{\lambda}}_{\text{CM}_3}\pd{\mathbf{K}}{\bm{\rho}}\mathbf{v}\\
&= \chi\left[\tr{\mathbf{u}}\pd{\mathbf{K}}{\bm{\rho}}\left(\frac{MSE}{(SE)^2}\left(-\frac{\mathbf{u}}{2}\right) + \frac{\mathbf{v}}{SE}\right)\right]\\&+ \underbrace{\chi\left[\left(\frac{MSE}{(SE)^2}\left(\tr{\mathbf{u}}\right) + \frac{-\tr{\mathbf{v}}}{SE}\right)\mathbf{T}\inv{\mathbf{A}}\pd{\mathbf{A}}{\bm{\rho}}\mathbf{p}\right]}_{\text{Load sensitivities}}
\end{split},
\end{equation}
Finally, one employs the chain rule in view with Eqs.~\eqref{eq:rho over psi} and \eqref{eq:objectivesensitivityrho} to determine derivatives of the objective functions with respect to design vector $\bm{\psi}$ as
\begin{equation}
\frac{\text{d}f_t^0}{\text{d}\bm{\psi}}|_{t= \text{s},\,\text{CM}} = \pd{f_t}{\bm{\rho}}\pd{\bm{\rho}}{\bm{\psi}}.
\end{equation}
and thus, the associated load sensitivities get evaluated computationally cheaply. Likewise, sensitivities of the constraints are determined.

\section{Numerical examples and discussion}\label{sec:NumericalExamples}
We solve design problems related to loadbearing structures (arch and piston) and CMs (inverter and gripper) involving pressure loads to demonstrate the versatility of the presented approach. The design domains with known boundary conditions for pressure loading and displacements are depicted in Figs.~\ref{fig:DesigndomainsStrucutres} and~\ref{fig:DesigndomainsCMs} for loadbearing structures and CMs, respectively. $\Gamma_\text{p}$ and $\Gamma_\mathrm{p_0}$ indicate the full and zero pressure loading boundaries, respectively. Optimization parameters and other specifications of the problems are tabulated in Table~\ref{Table:T1}, and any digression is reported in the associated problem definition. Implementation of the MMA with hexagonal FEs and elliptical masks is the same as the standard, except that after every MMA iteration, one determines the new mask vector/variable as
\begin{equation}\label{eq:MMAstep}
\bm{\psi}_\text{new} = \bm{\psi}_\text{old} + S\,\left(\bm{\psi}_\text{current} - \bm{\psi}_\text{old}\right),
\end{equation}
where $\bm{\psi}_\text{new}, \,  \bm{\psi}_\text{old}, \,\text{and}\, \bm{\psi}_\text{current}$ represent the new, old and current mask design variables. Note $\bm{\psi}_\text{current}$ is the solution obtained from the MMA optimizer using $\bm{\psi}_\text{old}$. $S$ indicates the length of a step one requires to multiply, which may depend upon the types of problems to be solved. In our experience, $S\in[0.01,\,0.1]$ can be a good choice for the used MMOS settings. For all the problems solved, dimensions in $x-$ and $y-$directions are denoted by $L_x$ and $L_y$ respectively. The number of FEs in $x-$ and $y-$directions are indicated by $N_\text{ex}$ and $N_\text{ey}$ respectively. That for masks are denoted by $N_\text{mx}$ and $N_\text{my}$ respectively. Thickness is set to $\SI{0.001}{\meter}$, and plane-stress conditions are assumed.

 \begin{figure*}
	\centering
	\begin{subfigure}{0.49\textwidth}
		\centering
		\includegraphics[scale=0.6]{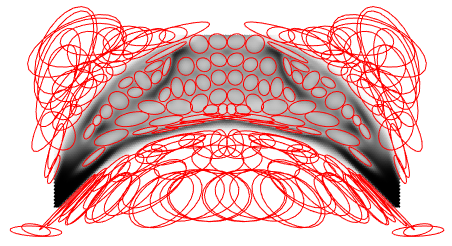}; 
		\caption{CASE~I: $\alpha_j = 1,\,\gamma_j=1$}
		\label{fig:lidalpetaone}
	\end{subfigure}
	\begin{subfigure}{0.49\textwidth}
		\centering
		\includegraphics[scale=0.54]{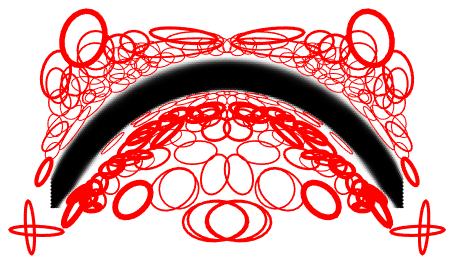} 
		\caption{CASE~II: $\alpha_j=1,\,\gamma_j$ as design variables}
		\label{fig:lidalpconsetaDV}
	\end{subfigure}
	\begin{subfigure}{0.47\textwidth}
		\centering
		\includegraphics[scale=0.6]{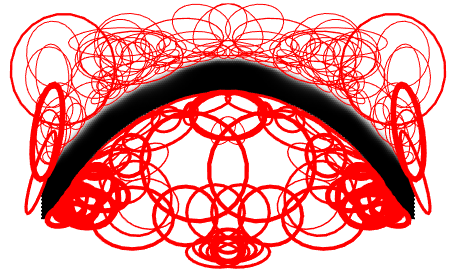}
		\caption{CASE~III: $\gamma_j = 1,\,\alpha_j$ as design variables}
		\label{fig:lidalpDVetacons}
	\end{subfigure}
	\quad
	\begin{subfigure}{0.47\textwidth}
		\centering
		\includegraphics[scale=0.6]{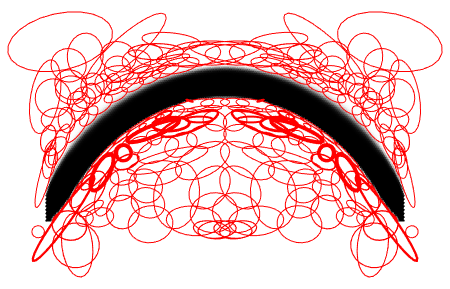}
		\caption{CASE~IV: $\alpha_j$ and $\gamma_j$ as design variables}
		\label{fig:lidalpDVetaDV}
	\end{subfigure}
	\caption{Results for four cases are displayed after 400 MMA iterations. (\subref{fig:lidalpetaone})  $f^0_s = \SI{0.81}{\newton\meter},\,V = 21\%,\,GS_\text{I} = 40.50\%$ (\subref{fig:lidalpconsetaDV})  $f^0_s = \SI{0.82}{\newton\meter},\, V = 20.4\%,\,GS_\text{I} = 5.91\%$ (\subref{fig:lidalpDVetacons}) $f^0_s = \SI{0.87}{\newton\meter},\,V = 18.2\%,\,GS_\text{I} = 5.3\%$ and (\subref{fig:lidalpDVetaDV})  $f^0_s = \SI{0.84}{\newton\meter},\,V=19.70\%,\,GS_\text{I} = 4.48\%$. $GS_\text{I}$ indicates gray scale indicator.}
	\label{fig:lidCaseStudy}
\end{figure*}
\begin{figure*}
	\centering
	\begin{subfigure}{0.475\textwidth}
		\centering
		\includegraphics[scale=0.45]{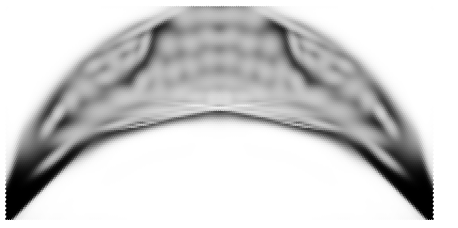} 
		\caption{CASE~I: $\alpha_j = 1,\,\gamma_j=1$}
		\label{fig:lidalpetaone_w_m}
	\end{subfigure}
	\begin{subfigure}{0.49\textwidth}
		\centering
		\includegraphics[scale=0.475]{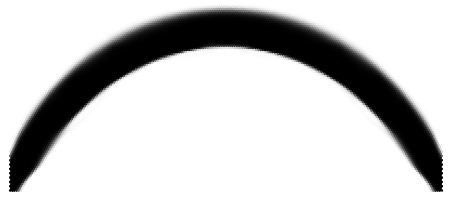} 
		\caption{CASE~II: $\alpha_j=1,\,\gamma_j$ as design variables}
		\label{fig:lidalpconsetaDV_w_m}
	\end{subfigure}
	\begin{subfigure}{0.47\textwidth}
		\centering
		\includegraphics[scale=0.475]{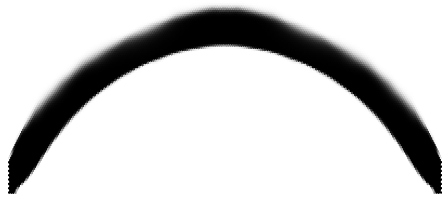}
		\caption{CASE~III: $\gamma_j = 1,\,\alpha_j$ as design variables}
		\label{fig:lidalpDVetacons_w_m}
	\end{subfigure}
	\begin{subfigure}{0.47\textwidth}
		\centering
		\includegraphics[scale=0.475]{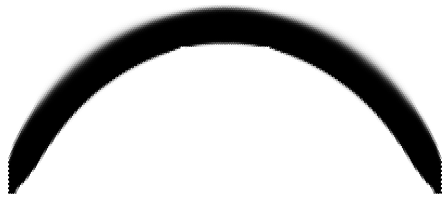} 
		\caption{CASE~IV: $\alpha_j$ and $\gamma_j$ as design variables}
		\label{fig:lidalpDVetaDV_w_m}
	\end{subfigure}
	\caption{Material distributions without masks of the four cases  shown in Fig.~\ref{fig:lidCaseStudy} are displayed.}
	\label{fig:lidCaseStudy_w_m}
\end{figure*}
\subsection{Internally pressurize arch}
TO problem for internally pressurized arch first presented in \cite{Hammer2000} is solved herein. The design specification is mentioned in Fig.~\ref{fig:lid}, and Table~\ref{Table:T1} indicates the design parameters employed. Area of the design domain is set to $L_x\times L_y$ = $0.2 \times 0.1$ \si{\square \meter}. The domain is parameterized using $N_\text{ex}\times N_\text{ey} = 200\times100$ hexagonal FEs using the $\texttt{HoneyMesher}$ code presented in~\cite{kumar2022honeytop90}. $N_\text{mx}\times N_\text{my} = m_n= 20\times 10$ elliptical masks are taken for optimization.

\subsubsection{Qualifying $\mathrm{\alpha_j}$ and $\mathrm{\gamma_j}$ as design variables}
Herein, a study is presented to indicate that indeed considering $\alpha_j$ and $\gamma_j$ as additional design variables can help achieve close to 0-1 optimized designs.

Four cases are conceptualized, CASE~I: $\alpha_j|_{j = 1,\,2,\,\cdots,\,m_n}= 1$ and $\gamma_j =1$, CASE~II: $\alpha_j = 1$ and $\gamma_j$ are included in the design variables with lower and upper bounds 1 and 30 respectively, CASE~III: $\gamma_j=1$ and $\alpha_j$ are included in the design variables with 1 and 30 as lower and upper bounds respectively, CASE~IV: $\alpha_j$ and $\gamma_j$ are considered design variables with bounds mentioned in CASE~II and CASE~III. Constraint $g_1$ is applied using $V^* = 0.20$. Step length is set to $S = 0.075$ (Eq.~\ref{eq:MMAstep}). 

\begin{figure*}[h!]
	\centering
	\begin{subfigure}[t]{0.45\textwidth}
		\centering
		\includegraphics[scale=0.5]{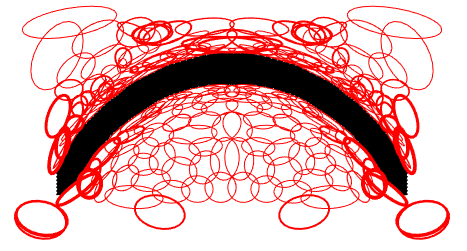}
		\caption{}
		\label{fig:archmasketa}
	\end{subfigure}
	\begin{subfigure}[t]{0.45\textwidth}
		\centering
		\includegraphics[scale=0.5]{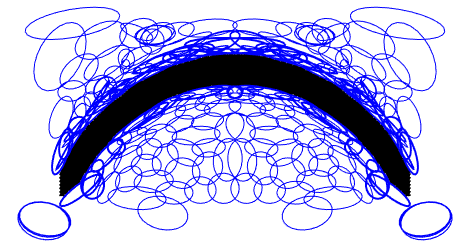}
		\caption{}
		\label{fig:archmaskalp}
	\end{subfigure}
	\quad
	\begin{subfigure}[t]{0.45\textwidth}
		\centering
		\includegraphics[scale=0.5]{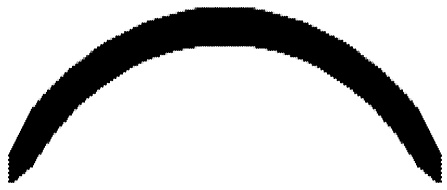}
		\caption{}
		\label{fig:archmaterial}
	\end{subfigure}
	\begin{subfigure}[t]{0.45\textwidth}
		\centering
		\includegraphics[scale=0.5]{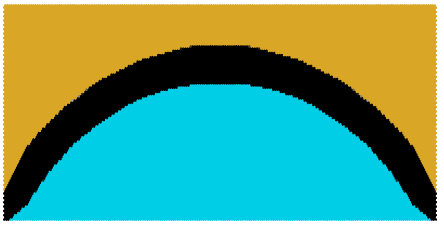}
		\caption{}
		\label{fig:archpressure}
	\end{subfigure}
	\caption{Optimized design for internally pressure loaded arch after 500 MMA iterations (\subref{fig:archmasketa}) Optimized material layout with final elliptical masks whose thickness are proportional to their $\gamma_j$, (\subref{fig:archmaskalp}) Material layout with final elliptical masks whose thickness are proportional to their $\alpha_j$, (\subref{fig:archmaterial}) Optimized material layout with $GS_\text{I} = 0.26\%$ and (\subref{fig:archpressure}) Optimized material layout with final pressure field.} \label{fig:archdesing}
\end{figure*}

Figure~\ref{fig:lidCaseStudy} depicts results for all the four cases with respective final compliance values, volume fractions and grayscale indicators $GS_\text{I}$. Results are displayed after 400 MMA iterations. The optimized design obtained for CASE~I has a relatively more number of gray elements, and therefore, it gets lower final strain energy than that of all other cases. One notices that CASE~IV indicates the lowest $GS_\text{I}$ value suggesting that the corresponding optimized design has lower gray FEs than others. In addition, the volume constraint gets satisfied. Further, $\alpha_j$ and $\gamma_j$ are treated as design variables, which in turn relatively enhances the search space in CASE~IV. In view of this study, we can conclude that considering  $\{\alpha_j,\,\gamma_j\}$ as design variables is indeed beneficial.

 Although using $\{\alpha_j,\,\gamma_j\}$ as design variables helps lower the number of gray FEs in the optimized designs (CASE~IV), one cannot explicitly control the grayscale indicator for the final solutions. To do that, we introduce grayscale indicator $GS_\text{I}$ constraint, i.e. $g_2$ constraint (Eq.~\ref{eq:Optimizationequation}) within the optimization formulation so that the selected discreteness level of the optimized designs can be achieved. Fig.~\ref{fig:lidCaseStudy_w_m} depicts the final designs without masks for four cases. Sharp corners are not seen, however, zigzag boundaries can be noted (Fig.~\ref{fig:lidCaseStudy_w_m}). This is because a set of finite elements constitutes the boundaries of the final designs in TO settings~\citep{sigmund2013topology,kumar2015topology}. Sec.~\ref{Sec:BS} shows that such undulations of the boundaries can be reduced by incorporating the boundary smoothing (BS) scheme~\citep{kumar2015topology} in the presented approach.

\subsubsection{Arch design}\label{sec:archdesign}

Having discussed using $\{\alpha_j,\,\gamma_j\}$ as design variables and  the requirement of $g_2$ constraint (Eq.~\ref{eq:Optimizationequation}), we solve the arch problem herein. We set the desired $GS_\text{I}$ to 0.3\%. The maximum number of MMA iterations is fixed to 500, and $S = 0.03$ is set (Eq.~\ref{eq:MMAstep}).

\begin{figure*}
	\centering
	\begin{subfigure}[t]{0.45\textwidth}
		\centering
		\begin{tikzpicture} 	
			\pgfplotsset{compat = 1.3}
			\begin{axis}[
				width = 1\textwidth,
				xlabel=MMA iterations,
				ylabel= $100 \times$ compliance ($\si{\newton\meter}$)]
				\pgfplotstableread{archobj.txt}\mydata;
				\addplot[mark=*,black]
				table {\mydata};
				\addlegendentry{$100 \times$Compliance}
			\end{axis}
		\end{tikzpicture}
		\caption{Objective convergence history}
		\label{fig:archobjhistory}
	\end{subfigure}
	\begin{subfigure}[t]{0.45\textwidth}
		\centering
		\begin{tikzpicture} 	
			\pgfplotsset{compat = 1.3}
			\begin{axis}[
				width = 1\textwidth,
				xlabel=MMA iterations,
				ylabel=  (\%)]
				\pgfplotstableread{archvolume.txt}\mydata;
				\addplot[mark=diamond*,black]
				table {\mydata};
				\addlegendentry{Volume fraction(\%)}
				\pgfplotstableread{archGI.txt}\mydata;
				\addplot[mark=otimes*,blue]
				table {\mydata};
				\addlegendentry{$GS_\text{I} (\%)$}
			\end{axis}
		\end{tikzpicture}
		\caption{Convergence plots for the constraints}
		\label{fig:archvolumeGIhistory}
	\end{subfigure}
	\caption{Objective and constraints history for the arch problem.}\label{fig:archobjvolumgrayscaleplots}
\end{figure*}
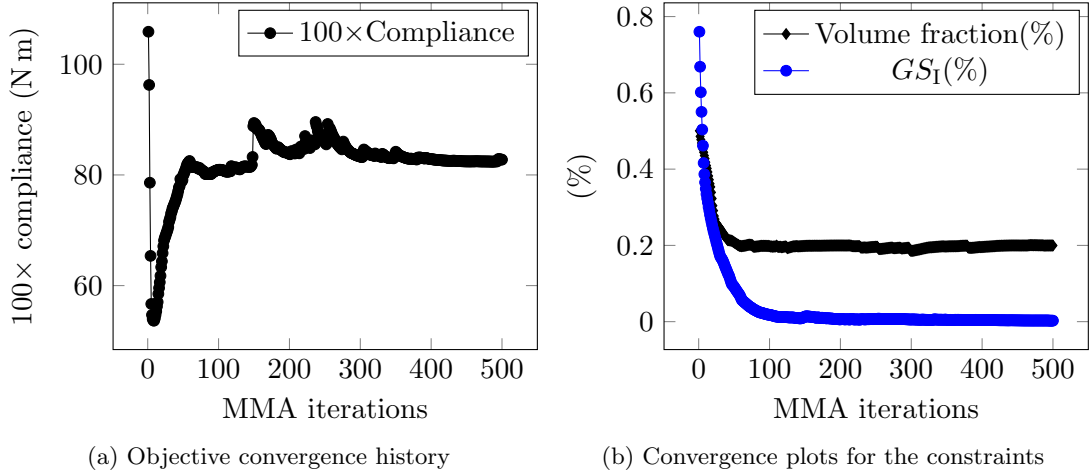

  Figures~\ref{fig:archdesing},~\ref{fig:archobjhistory} and~\ref{fig:archvolumeGIhistory} indicate the optimized designs and convergence history plots for objective, volume fraction  and grayscale indicator  after 500 MMA iterations respectively. The final shape, size, and orientation of masks are displayed with the optimized material layout in Fig.~\ref{fig:archmasketa} and~\ref{fig:archmaskalp} wherein the thickness of the masks boundaries are directly proportional to their $\gamma_j$ and $\alpha_j$, respectively. Masks with higher $\gamma_j$ can have lower $\alpha_j$ and vice versa.   The exclusive optimized material layout and that with pressure field are shown in Fig.~\ref{fig:archmaterial} and~\ref{fig:archpressure} respectively. Final optimized designs are similar to those obtained in \cite{Hammer2000,kumar2020topology}. The optimizer helps achieve the final design to contain the applied pressure loading with minimum compliance. The obtained final normalized compliance, volume fraction and grayscale indicator are \SI{0.83}{\newton \meter}, $0.20\%$ and $0.26\%$ respectively. The volume constraint is satisfied and active (Fig.~\ref{fig:archvolumeGIhistory}), whereas the grayscale constraint is satisfied at the end of the optimization. Thus the desired discreteness level is achieved.

\subsection{Piston design}
 The pressure loadbearing piston structure was first presented in \cite{bourdin2003design}, which is taken herein as a second structure problem. The design domain specification with dimension $L_x\times L_y = 0.12\times 0.04$ \si{\square\meter} is displayed in Fig.~\ref{fig:piston}. A vertical symmetry line exists for the design domain, which is used herein to solve only the symmetrical part of the domain. 

 We use $N_\text{ex}\times N_\text{ey} = 120\times80$ hexagonal FEs and  $N_\text{mx}\times N_\text{my} = 10\times 10$ elliptical negative masks to parametrize and  determine the optimized material layout of the symmetrical design, respectively. Volume fraction and grayscale constraint are set to 0.30 and 0.3\%, respectively. The upper bound on $\alpha_j$ is set to 40, and that on $\gamma_j$ is taken as 20. The maximum number of MMA iterations is set to 500. $S = 0.025$ is set for mask variable movement. $\{\beta_k,\,\beta_h\} = \{12,\,12\}$ and $\{\eta_k,\,\eta_h\} = \{0.25,\,0.25\}$ are considered. We refer to Table~\ref{Table:T1} for other design parameters.

\begin{figure*}[h!]
	\begin{subfigure}[t]{0.475\textwidth}
		\centering
		\includegraphics[scale=0.7]{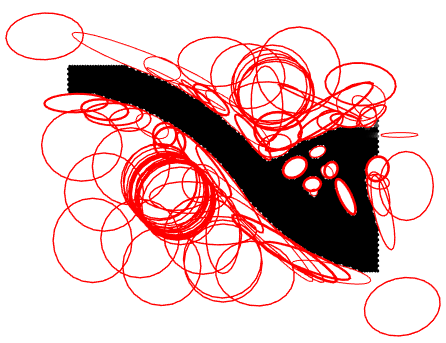} 
		\caption{}
		\label{fig:pistonmasketa}
	\end{subfigure}
	\quad
	\begin{subfigure}[t]{0.475\textwidth}
		\centering
		\includegraphics[scale=0.7]{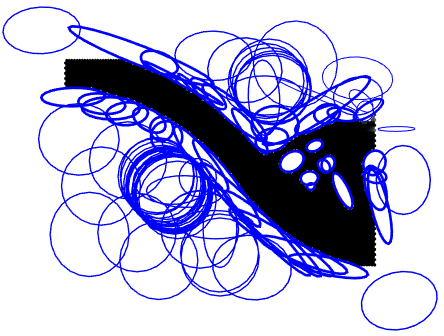}
		\caption{}
		\label{fig:pistonmaskalp}
	\end{subfigure}
	\quad
	\begin{subfigure}[t]{0.475\textwidth}
		\centering
		\includegraphics[scale=0.6]{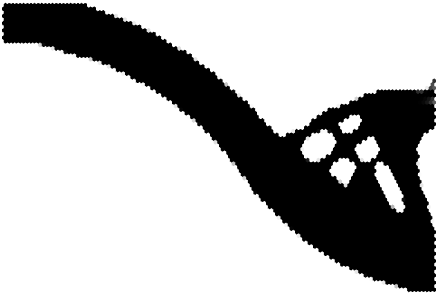}
		\caption{}
		\label{fig:pistonmaterial}
	\end{subfigure}
	\quad\quad\,
	\begin{subfigure}[t]{0.475\textwidth}
		\centering
		\includegraphics[scale=0.6]{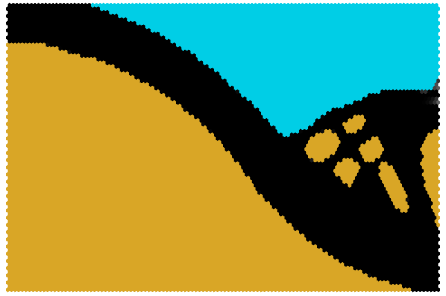}
		\caption{}
		\label{fig:pistonpressure}
	\end{subfigure}
	\caption{A symmetrical half optimized design for piston design after 500 MMA iterations (\subref{fig:pistonmasketa}) Optimized material layout with final elliptical masks whose line widths are proportional to their $\gamma_j$, (\subref{fig:pistonmaskalp}) Material layout with final elliptical masks whose line widths are proportional to their $\alpha_j$, (\subref{fig:pistonmaterial}) Optimized material layout with $GS_\text{I} = 0.3\%$ and (\subref{fig:pistonpressure}) Optimized material layout with final pressure field.} \label{fig:pistonoptimized}
\end{figure*}
\begin{figure*}
	\begin{subfigure}[t]{0.475\textwidth}
		\centering
\includegraphics[scale=0.6]{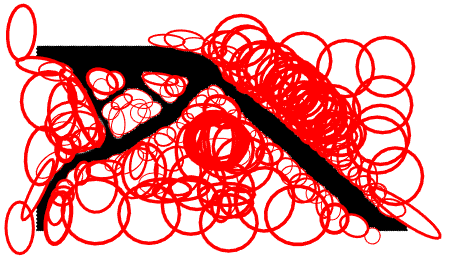}
		\caption{}
		\label{fig:IVCMmasketa}
	\end{subfigure}
	\begin{subfigure}[t]{0.475\textwidth}
		\centering
\includegraphics[scale=0.6]{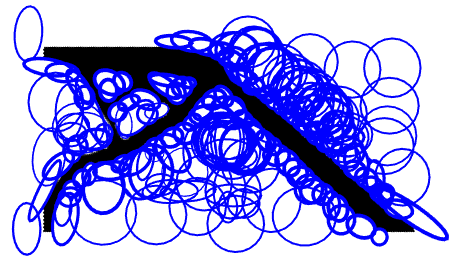}
		\caption{}
		\label{fig:IVCMmaskalp}
	\end{subfigure}
	\quad
	\begin{subfigure}[t]{0.475\textwidth}
		\centering
\includegraphics[scale=0.6]{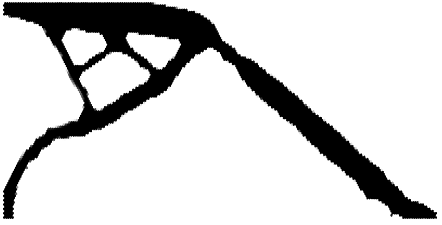}
		\caption{}
		\label{fig:IVCMmaterial}
	\end{subfigure}
	\begin{subfigure}[t]{0.475\textwidth}
		\centering
\includegraphics[scale=0.6]{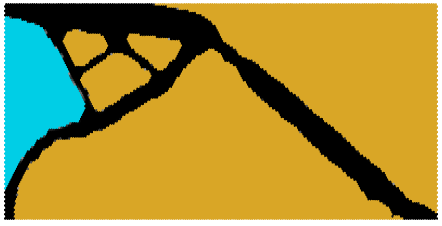}
		\caption{}
		\label{fig:IVCMpressure}
	\end{subfigure}
	\caption{A symmetrical half optimized design for inverter mechanism after 600 MMA iterations (\subref{fig:IVCMmasketa}) Optimized material layout with final elliptical masks whose line widths are proportional to their $\gamma_j$, (\subref{fig:IVCMmaskalp}) Material layout with final elliptical masks whose line widths are proportional to their $\alpha_j$, (\subref{fig:IVCMmaterial}) Optimized material layout with $GS_\text{I} = 0.5\%$ and (\subref{fig:IVCMpressure}) Optimized material layout with final pressure field.} \label{fig:IVCMptimized}
\end{figure*}

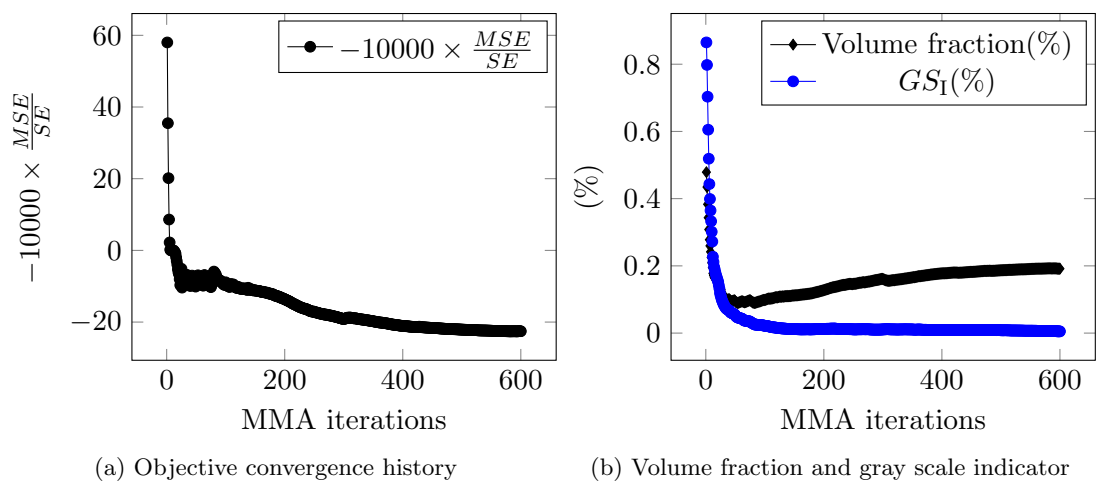
\begin{figure*}
	\centering
	\begin{subfigure}[t]{0.45\textwidth}
		\centering
		\begin{tikzpicture} 	
		\pgfplotsset{compat = 1.3}
		\begin{axis}[
		width = 1\textwidth,
		xlabel=MMA iterations,
		ylabel= $-10000\times\frac{MSE}{SE}$]
		\pgfplotstableread{IVCMobj.txt}\mydata;
		\addplot[mark=*,black]
		table {\mydata};
		\addlegendentry{$-10000\times\frac{MSE}{SE}$}
		\end{axis}
		\end{tikzpicture}
		\caption{Objective convergence history}
		\label{fig:IVCMobjhistory}
	\end{subfigure}
	\begin{subfigure}[t]{0.45\textwidth}
		\centering
		\begin{tikzpicture} 	
		\pgfplotsset{compat = 1.3}
		\begin{axis}[
		width = 1\textwidth,
		xlabel=MMA iterations,
		ylabel=  (\%)]
		\pgfplotstableread{IVCMvolume.txt}\mydata;
		\addplot[mark=diamond*,black]
		table {\mydata};
		\addlegendentry{Volume fraction(\%)}
		\pgfplotstableread{IVCMGI.txt}\mydata;
		\addplot[mark=otimes*,blue]
		table {\mydata};
		\addlegendentry{$GS_\text{I} (\%)$}
		\end{axis}
		\end{tikzpicture}
		\caption{Volume fraction and gray scale indicator}
		\label{fig:IVCMvolumeGIhistory}
	\end{subfigure}
	\caption{Objective and constraints convergence plots for the inverter mechanism.}\label{fig:IVCMobjvolumgrayscaleplots}
\end{figure*}

\begin{figure*}
	\begin{subfigure}[t]{0.5\textwidth}
		\centering
\includegraphics[scale=0.62]{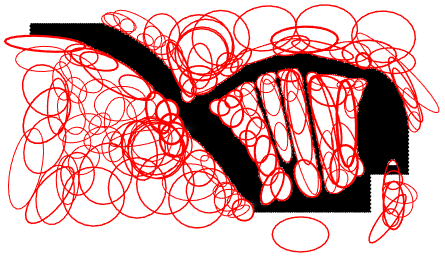}
		\caption{}
		\label{fig:Grippermasketa}
	\end{subfigure}
	\begin{subfigure}[t]{0.5\textwidth}
		\centering
\includegraphics[scale=0.6]{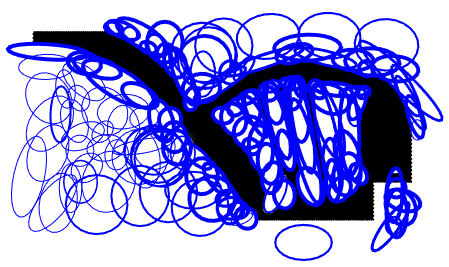}
		\caption{}
		\label{fig:Grippermaskalp}
	\end{subfigure}
	\quad
	\begin{subfigure}[t]{0.5\textwidth}
		\centering
\includegraphics[scale=0.6]{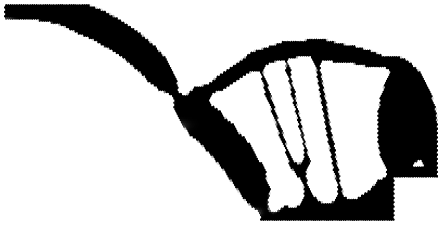}
		\caption{}
		\label{fig:Grippermaterial}
	\end{subfigure}
	\begin{subfigure}[t]{0.5\textwidth}
		\centering
\includegraphics[scale=0.6]{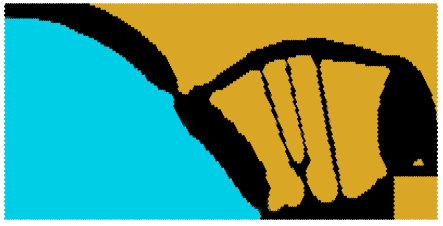}
		\caption{}
		\label{fig:Gripperpressure}
	\end{subfigure}
	\caption{A symmetrical half optimized design for gripper mechanism after 600 MMA iterations (\subref{fig:Grippermasketa}) Optimized material layout with final elliptical masks whose line widths are proportional to their $\gamma_j$, (\subref{fig:Grippermaskalp}) Material layout with final elliptical masks whose line widths are proportional to their $\alpha_j$, (\subref{fig:Grippermaterial}) Optimized material layout with $GS_\text{I} = 0.6\%$, and (\subref{fig:Gripperpressure}) Optimized material layout with final pressure field.} \label{fig:Gripperoptimized}
\end{figure*}

 A symmetrical half-optimized piston design is displayed in Fig.~\ref{fig:pistonoptimized}. Plots with masks considering the values of $\gamma_j$ and $\alpha_j$ proportion to the line widths of masks are depicted in Fig.~\ref{fig:pistonmasketa} and~\ref{fig:pistonmaskalp} respectively. One notices that $\alpha_j$ and $\gamma_j$ vary differently, as noted in the arch problem result. Specifically, nearly all masks whose boundaries define the contour of the continuum seem to have higher $\alpha_j$, as expected, since 
this helps boundary FEs attain states close to the solid state. However, not all $\gamma_j$ are high for the same masks. This suggests that
selective/local dilation/erosion may occur at the continuum boundaries to satisfy the grayscale constraint. The optimized piston design resembles the previously obtained results for the same problem~\citep{bourdin2003design,kumar2020topology}. The final compliance, volume fraction and $GS_\text{I}$ are $\SI{10.97}{\newton\meter}$, $30\%$ and $0.3\%$, respectively. The volume constraint and grayscale constraint are satisfied and active at the end of optimization, indicating that the desired discreteness level is reached. Next, we solve pressure-actuated compliant mechanisms.

\subsection{Pressure-driven CMs}\label{Sec:CMs}
 Pressure-actuated inverter and Gripper CMs are designed using a multi-criterion objective (Eq.~\ref{eq:Optimizationequation}) with volume and grayscale indicator constraints.

The symmetric half designs for inverter and gripper mechanisms are depicted in Fig.~\ref{fig:inverter} and~\ref{fig:gripper}, respectively. $L_x\times L_y$ = $0.2 \times 0.1$ \si{\square \meter} is set for each mechanism. Figs.~\ref{fig:inverter} and~\ref{fig:gripper} also depict each mechanism's output location and direction of movement using thick red arrows. For the inverter mechanism, an inverse motion with respect to the pressure loading direction is sought, whereas a perpendicular gripping motion is desired in the case of the gripper mechanism. To provide a proper seat for the workpiece, a void passive region having dimension $\frac{L_x}{10}\times \frac{L_x}{10}$ is provided and for gripping jaws (solid passive regions) dimension $\frac{L_x}{10}\times\frac{L_x}{40}$ are set. Springs with stiffness $k_\text{s} = \SI{1e4}{\newton\meter}$ and $\SI{5e4}{\newton\meter}$ are attached at the output location of inverter and gripper mechanisms, respectively. These springs represent the workpiece stiffnesses at the output locations.  $N_\text{ex}\times N_\text{ey} = 200 \times 100$ FEs are employed to describe the design domains. The number of elliptical masks is set to $N_\text{mx}\times N_\text{my}$ = $20\times 10$ for the inverter and $N_\text{mx}\times N_\text{my}$ = $12\times 12$ for the gripper mechanisms are employed. Volume fractions for inverter and gripper mechanisms are set to 0.20 and 0.25, respectively. $\delta$ for $GS_\text{I}$ is chosen to $0.005$ for both the mechanisms.  The maximum number of MMA iterations is set to 600. The upper bounds on $\alpha_j$ and $\gamma_j$ for the inverter mechanism used are 60 and 50, while for the gripper mechanisms, those are 60 and 20, respectively. $\{\beta_k,\,\beta_h\} = \{12,\,12\}$ and $\{\eta_k,\,\eta_h\} = \{0.25,\,0.25\}$ are considered for the inverter mechanism, whereas those for the gripper mechanism are $\{10,\,10\}$ and $\{\eta_k,\,\eta_h\} = \{0.25,\,0.25\}$. For both mechanisms, step lengths (Eq.~\ref{eq:MMAstep}) is set to 0.01.

\begin{figure*}
	\begin{subfigure}[t]{0.45\textwidth}
		\centering
\includegraphics[scale=0.56]{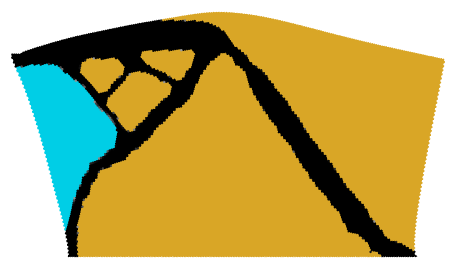}
		\caption{}
		\label{fig:DeformedIV}
	\end{subfigure}
	\quad\,\,
	\begin{subfigure}[t]{0.45\textwidth}
		\centering
\includegraphics[scale=0.6]{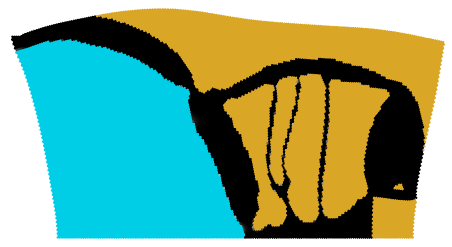}
		\caption{}
		\label{fig:DeformedGP}
	\end{subfigure}
	\caption{Deformed profiles for inverter and gripper mechanisms are displayed in (\subref{fig:DeformedIV}) and (\subref{fig:DeformedGP}) respectively. } \label{fig:DeformedIVGP}
\end{figure*}

\begin{figure*}[h!]
	\centering
	\begin{subfigure}{0.45\textwidth}
		\centering
		\includegraphics[scale=0.5]{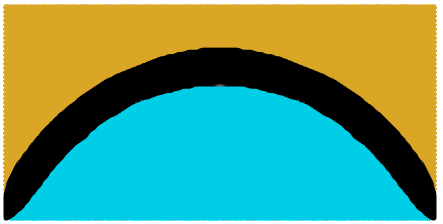} 
		\caption{$\beta = 8$}
		\label{fig:smootharch}
	\end{subfigure}
	\begin{subfigure}{0.45\textwidth}
		\centering
		\includegraphics[scale=0.38]{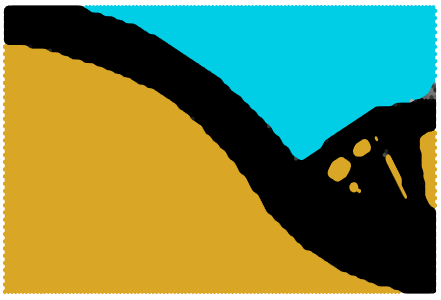} 
		\caption{$\beta = 4$}
		\label{fig:smoothpiston}
	\end{subfigure}
	\begin{subfigure}{0.45\textwidth}
		\centering
		\includegraphics[scale=0.5]{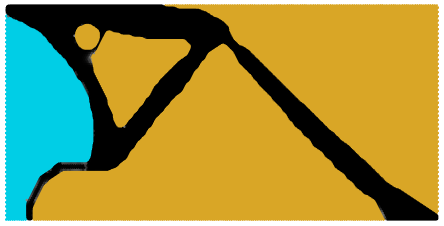}
		\caption{$\beta = 6$}
		\label{fig:smoothinverter}
	\end{subfigure}
	\begin{subfigure}{0.45\textwidth}
		\centering
		\includegraphics[scale=0.5]{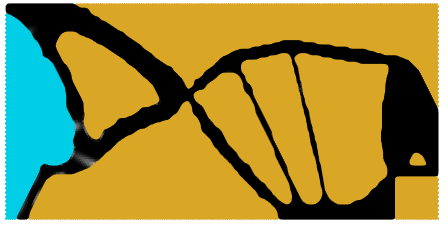} 
		\caption{$\beta= 6$}
		\label{fig:smoothgripper}
	\end{subfigure}
	\caption{Smooth optimized designs. (\subref{fig:smootharch}) Optimized smooth arch, (\subref{fig:smoothpiston}) Symmetric half optimized  smooth piston, (\subref{fig:smoothinverter}) Symmetric half optimized  smooth inverter, and (\subref{fig:smoothgripper}) Symmetric half optimized  smooth gripper.}
	\label{fig:smoothresults}
\end{figure*}

The optimized designs for inverter and gripper mechanisms are depicted in Fig.~\ref{fig:IVCMptimized} and \ref{fig:Gripperoptimized} respectively. Masks with optimized CMs with $\gamma_j$ and $\alpha_j$ represented by their line thickness are plotted in Fig.~\ref{fig:IVCMmasketa}, \ref{fig:Grippermasketa} and Fig.~\ref{fig:IVCMmaskalp}, \ref{fig:Grippermaskalp},  respectively.  Fig.~\ref{fig:IVCMmaterial} and \ref{fig:Grippermaterial} show the optimized results with pressure field. The final volume fraction for inverter and gripper mechanisms are 0.20 and 0.25, respectively, and the final recorded grayscale indicator are 0.5\% and 0.6\%, respectively. Volume constraints are satisfied and active at the end of the optimization for both cases. The desired discreteness level is achieved in the case of the inverter mechanism, whereas for the gripper mechanism, the achieved discreteness level is close to the desired part. This may be because the material density of each FE is a cumulative effect of all mask shapes, sizes, positions, and orientations (see Eq.~\ref{eq:materialdensityofeachFE1}). After a limit for a given problem setting with $GS_\text{I}$ constraint, it may be difficult for the optimizer to move toward a better solution. Nevertheless, it can be inferred that the lower grayscale indicator constraint while keeping $\alpha_j$ and $\gamma_j$ as additional design variables helps in achieving close to 0-1 solutions. Furthermore, in certain cases, one may achieve the target $\delta$ (or lower than that) by providing a range of  $\alpha_j$ and $\gamma_j$ high, e.g., [1,\, 200]. However, such limits could potentially jeopardize the optimization process by providing sensitivities close to zero. This can be one of the limitations of the proposed method; however, this is very much in line with the gradient-based TO, wherein some FEs, especially at boundaries,  will have gray nature. The optimized CMs with pressure fields for the inverter and gripper mechanisms are depicted in Fig.~\ref{fig:IVCMpressure} and~\ref{fig:Gripperpressure}, respectively. One notes that to contain the pressure loads, the optimizer provides a chamber-like inflated design at the input locations. The objective and constraints convergence plots for the optimized inverter mechanism are displayed in Fig.~\ref{fig:IVCMobjhistory} and~\ref{fig:IVCMvolumeGIhistory} respectively. At the end of the optimization process, these plots converge smoothly. One can note that not all masks with high $\gamma_j$ are the same as those with high $\alpha_j$, and vice versa. The dilation and erosion variables for each mask get  selective optimal values so that number of grey cells at the boundaries are minimized overall, e.g., Figs.~\ref{fig:archmasketa}-\ref{fig:archmaskalp}, \ref{fig:pistonmasketa}-\ref{fig:pistonmaskalp}, \ref{fig:IVCMmasketa}-\ref{fig:IVCMmaskalp} and \ref{fig:Grippermasketa}-\ref{fig:Grippermaskalp}. The deformed profiles for the inverter and gripper mechanisms with their pressure field are illustrated in Fig.~\ref{fig:DeformedIV} and~\ref{fig:DeformedGP} respectively. The obtained motions of the output nodes of the mechanisms are as they are designed for. Next, we solve all the examples using the boundary smoothing scheme within the approach as per~\cite{kumar2015topology}.

\begin{figure*}
	\centering
	\begin{subfigure}{0.45\textwidth}
		\centering
		\begin{tikzpicture} 	
			\pgfplotsset{compat = 1.3}
			\begin{axis}[
				width = 1\textwidth,
				xlabel=MMA iterations,
				ylabel= $100 \times$ compliance ($\si{\newton\meter}$)]
				\pgfplotstableread{archsmoothobj.txt}\mydata;
				\addplot[mark=*,black]
				table {\mydata};
				\addlegendentry{$100 \times$Compliance}
			\end{axis}
		\end{tikzpicture}
		\caption{Objective convergence history}
		\label{fig:archsmoothobjhistory}
	\end{subfigure}
	\begin{subfigure}{0.45\textwidth}
		\centering
		\begin{tikzpicture} 	
			\pgfplotsset{compat = 1.3}
			\begin{axis}[
				width = 1\textwidth,
				xlabel=MMA iterations,
				ylabel=  (\%)]
				\pgfplotstableread{archsmoothvolume.txt}\mydata;
				\addplot[mark=diamond*,black]
				table {\mydata};
				\addlegendentry{Volume fraction(\%)}
				\pgfplotstableread{archsmoothGI.txt}\mydata;
				\addplot[mark=otimes*,blue]
				table {\mydata};
				\addlegendentry{$GS_\text{I} (\%)$}
			\end{axis}
		\end{tikzpicture}
		\caption{Convergence plots for the constraints}
		\label{fig:archsmoothvolumeGIhistory}
	\end{subfigure}
	\caption{Objective and constraints history for the arch problem with the BS technique.}\label{fig:archsmoothobjvolumgrayscaleplots}
\end{figure*}
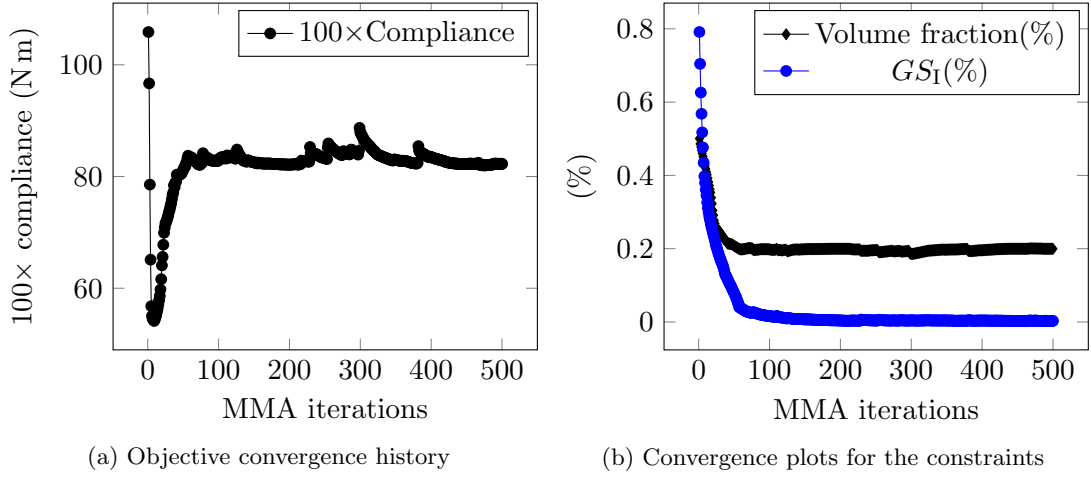
\begin{figure*}
	\centering
	\begin{subfigure}{0.45\textwidth}
		\centering
		\begin{tikzpicture} 	
			\pgfplotsset{compat = 1.3}
			\begin{axis}[
				width = 1\textwidth,
				xlabel=MMA iterations,
				ylabel= $-10000\times\frac{MSE}{SE}$]
				\pgfplotstableread{IVCMsmoothobj.txt}\mydata;
				\addplot[mark=*,black]
				table {\mydata};
				\addlegendentry{$-10000\times\frac{MSE}{SE}$}
			\end{axis}
		\end{tikzpicture}
		\caption{Objective convergence history}
		\label{fig:IVCMsmoothobjhistory}
	\end{subfigure}
	\begin{subfigure}{0.45\textwidth}
		\centering
		\begin{tikzpicture} 	
			\pgfplotsset{compat = 1.3}
			\begin{axis}[
				width = 1\textwidth,
				xlabel=MMA iterations,
				ylabel=  (\%)]
				\pgfplotstableread{IVCMsmoothvolume.txt}\mydata;
				\addplot[mark=diamond*,black]
				table {\mydata};
				\addlegendentry{Volume fraction(\%)}
				\pgfplotstableread{IVCMsmoothGI.txt}\mydata;
				\addplot[mark=otimes*,blue]
				table {\mydata};
				\addlegendentry{$GS_\text{I} (\%)$}
			\end{axis}
		\end{tikzpicture}
		\caption{Volume fraction and gray scale indicator}
		\label{fig:IVCMsmoothvolumeGIhistory}
	\end{subfigure}
	\caption{Objective and constraints convergence plots for the inverter mechanism with the BS scheme.}\label{fig:IVCMsmoothobjvolumgrayscaleplots}
\end{figure*}
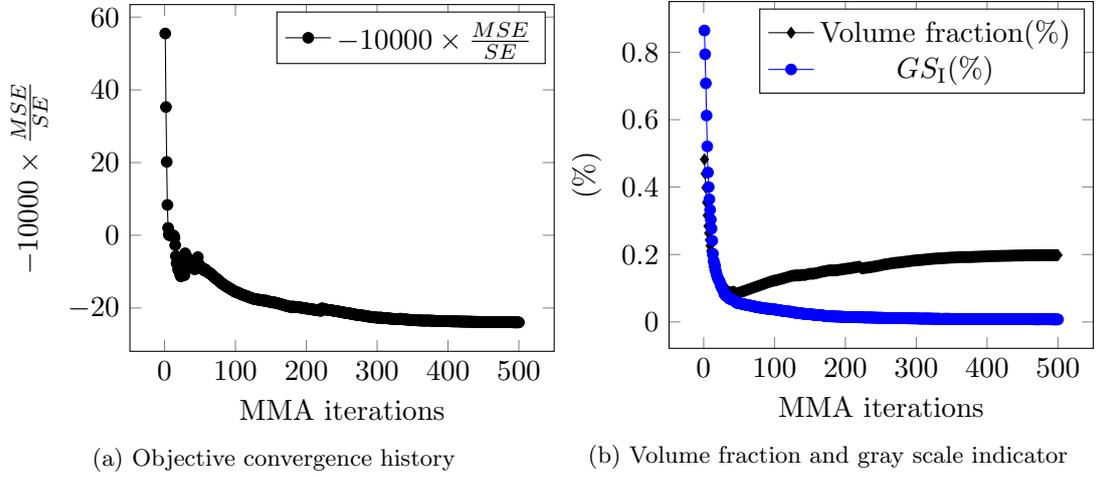
\subsection{Results with the boundary smoothing scheme}\label{Sec:BS}

One can note that although the obtained final designs of the loadbearing arch (Fig.~\ref{fig:archdesing}) and piston (Fig.~\ref{fig:pistonoptimized}), and pressure-actuated inverter (Fig.~\ref{fig:IVCMptimized}) and gripper (Fig.~\ref{fig:Gripperoptimized}) mechanisms are close to 0-1 solutions, their boundaries  contain V-notches\footnote{Optimized designs obtained with rectangular discretization contain right-angled notches} that pose challenges in manufacturing. Therefore, to suppress such notches, we have used the BS technique per \cite{kumar2015topology} within the proposed approach. The scheme determines the boundary nodes and shifts them systematically \citep{kumar2015topology}. Midpoints of the boundaries are connected via straight lines, and boundary nodes are then projected onto the lines along their shortest perpendiculars. The smoothing step can be performed $\beta \ge 1$ (integer) times. New positions of the nodes are used within the optimization steps while retaining the connectivity matrix and non-boundary nodes.

Figure~\ref{fig:smoothresults} shows the optimized results with the BB scheme~\citep{kumar2015topology}. $\beta=8$ and $\beta=4$ are used to obtain the loadbearing arch and piston structures, respectively. For inverter and gripper mechanisms, $\beta = 6$ is used. One can note that the boundaries of the results obtained with the BS scheme are relatively smoother than their counterparts solved without the smoothing scheme and have different topologies. Fig.~\ref{fig:archsmoothobjvolumgrayscaleplots} and Fig.~\ref{fig:IVCMsmoothobjvolumgrayscaleplots} depict  convergence curves for the internally pressurized arch and pressure-driven inverter mechanism. One can note that at the end of the optimization iterations, these plots converge smoothly.

\section{Closure}\label{sec:closure}

The presented MMOS topology optimization approach gives pressure-loaded structure and pressure-actuated compliant mechanism designs close to the desired discreteness level. The final performances of these mechanisms are as expected. Negative elliptical masks are used, and for each mask, in addition to its position, size, and orientation, the logistic variable (material dilation) and exponent (material erosion) are posed as design variables. A high value of the logistic variable leads to material addition near the mask boundary. In contrast, the significant value of the exponent results in material erosion inside and outside the boundary. By optimally determining their values for each mask, finite element densities can be controlled indirectly, leading to nearly black and white topologies. An explicit grayscale constraint is employed that helps achieve the desired discreteness level of the optimized topologies. The boundary smoothing scheme is used within the proposed approach. The results obtained using the smoothing scheme have relatively smooth boundaries and thicker members (for the gripper mechanism). The objective and constraints history curves converge and smooth at the end of the optimization iterations.

Hexagonal elements (honeycomb tessellation) describe the design domains that provide edge connectivity; thus, point connections and checkerboard patterns automatically vanish from the optimized designs. Darcy's law with a drainage term is employed to relate the pressure field with the material density vector, wherein the flow coefficient of each element is interpolated using a smooth Heaviside projection function in line with~\citet{kumar2020topology}. The formulation implicitly facilitates determining pressure loading surfaces/curves as the topology optimization evolves, wherein the span of pressure gradient alters with topology optimization iterations. The pressure field is then transformed into nodal forces using Wachspress shape functions employed to model hexagonal elements. The importance of drainage term with hexagonal elements is also demonstrated using a design domain containing multiple solid finite elements layers. The approach provides an automatic and computationally inexpensive evaluation of the load sensitivity terms while determining objective sensitivities using the adjoint-variable method in association with the chain rule.  

The optimized pressure-actuated compliant mechanisms are designed with small deformation mechanics assumptions. The obtained output performance of these mechanisms as they are sought for. Extending the approach for finite deformation problems for soft (compliant) robotic designs will have additional challenges, e.g., treating the pressure loads as follower forces. Thus, it needs a dedicated and detailed investigation, which can be one of the engaging future directions. Extending the proposed methodology to three dimensions with spheroidal masks can be another prospective study.  
	\section*{Acknowledgment}
	P. Kumar acknowledges financial support from the Science \& Engineering research board, Department of Science and Technology, Government of India under the project file number RJF/2020/000023. A. Saxena acknowledges the Alexander von Humboldt Foundation, IGMR, RWTH-Aachen, and 
	Politechnika Gda\'{n}ska for all their support. The authors thank Prof. Krister Svanberg for providing MATLAB codes of the MMA optimizer. 
	\bibliography{Referencesfinal}
   \bibliographystyle{spbasic} 
\end{document}